\documentclass[prb,amsfonts,amssymb,twocolumn,floats,superscriptaddress,aps]{revtex4-2}
\usepackage{setspace}
\usepackage[latin1]{inputenc}
\usepackage{graphicx}
\usepackage{physics}
\usepackage{amssymb,amsmath,braket}
\usepackage{upgreek}
\usepackage{bm,bbold}
\usepackage[dvipsnames]{xcolor}
\usepackage{graphicx,tikz}
\usepackage[normalem]{ulem}
\usepackage{color, colortbl}
\usepackage{mathbbol}
\usepackage{bbm}
\usepackage{tikz}

\usepackage{hhline}
\usepackage{array}
\usepackage{float}
\usepackage{subfigure}
\usepackage{multirow}

\usepackage[colorlinks,linktocpage]{hyperref}
\hypersetup{
	citecolor  = NavyBlue,
	linkcolor  = NavyBlue,
	urlcolor = NavyBlue
}

\DeclareMathOperator{\sgn}{sgn}

\DeclareMathOperator{\im}{Im}

\newcommand{\iu}{\mathrm{i}}
\newcommand{\eu}{\mathrm{e}}

\newcommand{\mac}{\mathcal}

\newcommand{\be}{\begin{equation}}
\newcommand{\ee}{\end{equation}}
\newcommand{\beq}{\begin{equation}}
\newcommand{\eeq}{\end{equation}}

\newcommand*\diff{\mathop{}\!\mathrm{d}}

\newcommand{\non}{\nonumber}
\newcommand{\dg}{\dagger}
\newcommand{\ve}[1]{{\bm #1}}

\newcommand{\kv}{\ve{k}}
\newcommand{\qv}{\ve{q}}
\newcommand{\Qv}{\ve{Q}}

\newcommand{\Dpg}{\Delta_{\rm pg}}
\newcommand{\tDpg}{\Delta^{(1)}_{\rm pg}}
\newcommand{\eV}{\text{eV}}
\newcommand{\sump}{\sideset{}{'}\sum}
\newcommand{\bw}{\bar{\omega}}
\newcommand{\bbt}{\mathbb{t}}

\newcommand{\e}{\epsilon}

\newcommand{\emi}{\varepsilon}

\definecolor{bananayellow}{rgb}{1.0, 0.88, 0.21}
\definecolor{straw}{rgb}{0.32, 0.28, 0.1}

\begin{document}

\title{Location and thermal evolution of the pseudogap due to spin fluctuations}
\author{Mengxing Ye}
\affiliation{Kavli Institute for Theoretical Physics, University of California, Santa Barbara, CA 93106, USA}
\affiliation{Department of Physics and Astronomy, University of Utah, Salt Lake City, UT 84112, USA}
\author{Zhentao Wang}
\thanks{Present address: Center for Correlated Matter and School of Physics, Zhejiang University, Hangzhou 310058,
China}
\affiliation{School of Physics and Astronomy, University of Minnesota, Minneapolis, MN 55455, USA}
\author{Rafael M Fernandes}
\affiliation{School of Physics and Astronomy, University of Minnesota, Minneapolis, MN 55455, USA}
\author{Andrey V Chubukov}
\affiliation{School of Physics and Astronomy and William I. Fine Theoretical Physics Institute, University of
Minnesota, Minneapolis, MN 55455, USA}
\begin{abstract}
We study pseudogap behavior in a metal near a spin density wave (SDW) instability due to thermal magnetic fluctuations. We consider the $t-t'$ Hubbard model on a square lattice at a finite doping, at intermediate coupling strength, and analyze the thermal evolution of the electron spectral function between a SDW ordered state at low temperatures and a normal Fermi liquid at high temperatures. We argue that for proper description of the pseudogap one needs to sum up infinite series of diagrams  for both the fermionic self-energy and the SDW order parameter in the SDW state or the magnetic correlation length in the paramagnetic state.
We use the eikonal approach to sum up an infinite series of diagrammatic contributions from thermal fluctuations.
Earlier studies found that in the SDW state, the spectral function $A_\kv (\omega)$ of a hot fermion  at a finite $T$ is exponentially small below the energy scale $\Delta (T)$, which scales with SDW order and vanishes at the ordering temperature $T_N$, and has a  hump at a larger frequency $\Dpg$, comparable to the zero-temperature SDW gap $\Delta (T=0)$. We argue that the hump, which we associate with the pseudogap, survives in some $T$ range above $T_N$. We show that this range is split by regions of strong and weak pseudogap behavior. In the first region, $\Dpg$ is weakly temperature dependent, despite that it comes from thermal fluctuations.  Such a behavior has been seen in numerical studies of the Hubbard model.  We show that  to obtain it, one needs to go beyond the one-loop approximation and sum up the infinite series of diagrams. In the second regime, $\Dpg$ decreases with increasing $T$ and eventually vanishes. We further argue that a magnetic pseudogap at a finite $T$  emerges only if the ground state is magnetically ordered. We present the phase diagram and apply the results to high-$T_c$ cuprates.
 \end{abstract}
\date{\today}
\maketitle

\section{Introduction}

The origin of the pseudogap behavior, observed in the cuprates and other correlated materials,  is still a subject of
ongoing debates.
Theoretical proposals for the pseudogap can be broadly split into three categories. One set of proposals is that the
pseudogap phase is a new state of matter with some particle-hole order. The order can be either a conventional one,
like  spin-density wave (SDW) or charge-density wave
(CDW)~\cite{Metlitski2010,WangYuxuan2014,Chowdhury2014,Atkinson2015,Grilli}, or less conventional, like a circulating
current~\cite{Varma1997,Varma1999}. The second type of proposals is that the pseudogap phase is a state with a
topological order, whose feedback effect on fermions mimics that of a SDW
order~\cite{Sachdev2019,YHZhang2020a,Sachdev2022pga,Sachdev2022pgb,Sachdev202302}. Finally, the third set of proposasl
is that pseudogap is not an ordered state, but rather a precursor to either a spin-density-wave (SDW) order~\cite{Vilk1996,Vilk1997,Schmalian1998,Schmalian1999,Sadovskii1999,Moca2000,Sadovskii_review,Yanase2004,Roy_2008,Sedrakyan2010,Gull2015,Gunnarsson2015,
Ye2019pg,Schafer2021,Held2022,Krien2021,Simkovic2022,*Simkovic2022a,Ye2023b},  or
superconductivity~\cite{Randeria1998,Millis1998,Fujimoto2002,Yanase2004,Berg2007,YMWu2021,Dai2021,Qi2022},
 or pair-density-wave~\cite{Dai2020}.

This paper is devoted to the analysis of the third scenario, more specifically to
 precursors to
 $(\pi, \pi)$ antiferromagnetic order in 2D.
The generic motivation here is based on neutron scattering, x-ray, and other measurements, which show that,
 e.g.,  in the cuprates,
 magnetic fluctuations remain strong in the paramagnetic phase
 in a sizable range of dopings and temperatures, which includes the pseudogap region (see
 e.g.~\cite{DamascelliRMP,ArmitageRMP,NormalPseudogapReivew2005} and references therein).
We note in passing that the pairing interaction, mediated by soft overdamped spin fluctuations, is attractive in the
$d$-wave channel; as such, a spin-fluctuation scenario for pairing has been widely discussed for cuprates and other
materials~\cite{ScalapinoRMP}.

\begin{figure*}[t]
\subfigure[]{\includegraphics[width=0.55\columnwidth]{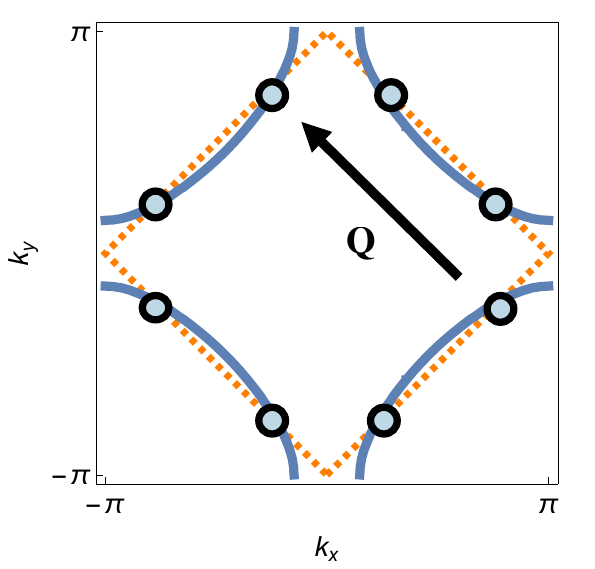}}
\subfigure[]{\includegraphics[width=0.7\columnwidth]{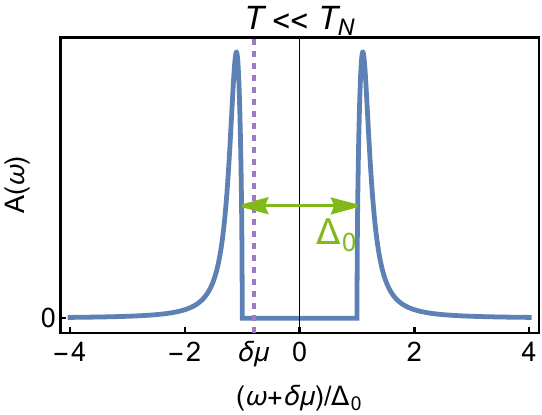}}
\subfigure[]{\includegraphics[width=0.7\columnwidth]{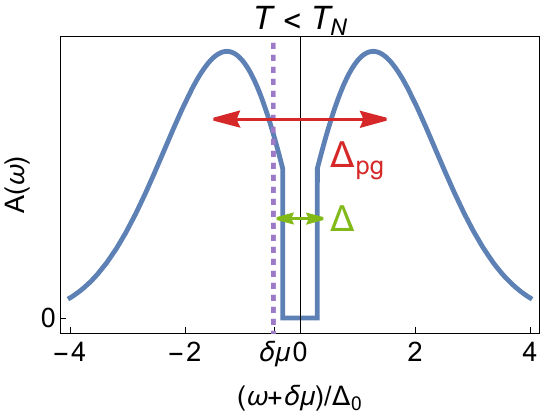}}
\subfigure[]{\includegraphics[width=0.7\columnwidth]{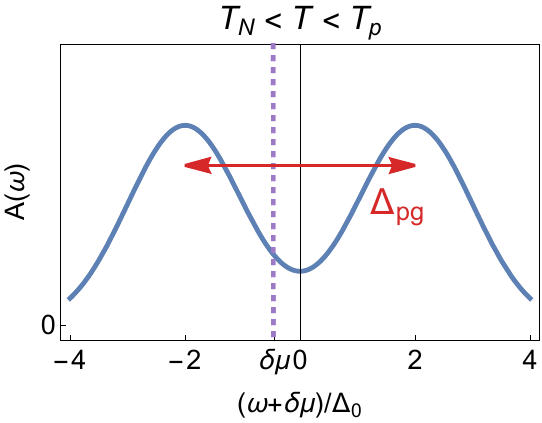}}
\subfigure[]{\includegraphics[width=0.7\columnwidth]{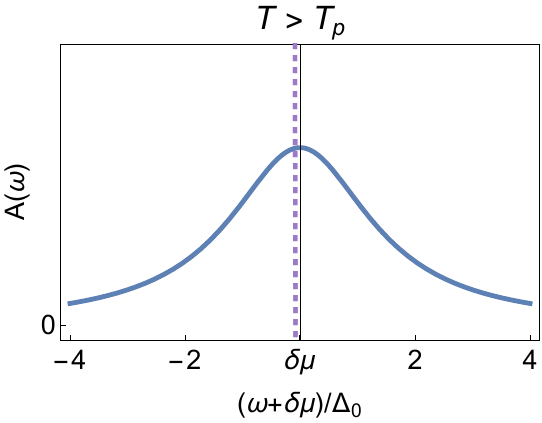}}
\caption{
(a) Fermi surface  of the $t-t'$ model on a square lattice.  The eight blue spots indicate the ``hot spots"  that
satisfy $\epsilon_\kv = \epsilon_{\kv + \Qv}$, where $\Qv=(\pi,\pi)$.
The orange dashed line specifies the folded Brillouin zone in the presence of $(\pi,\pi)$ order.
(b)-(e): Schematic plots of the electron spectral function at the hot spot (b) in the SDW state for $T\ll T_N$, (c)
in the SDW state at $T$ close to $T_N$, (d) in the pseudogap metal state, (e) in the normal Fermi liquid state.
$\Delta_0\approx U/2$
  in panel (b)
is the mean field SDW order parameter at $T=0$.
The quantity $\delta \mu = \mu - \mu_0$, shown by a purple dashed line, is the difference between the actual chemical potential $\mu$ and the
chemical potential for free fermions $\mu_0$.
Determining temperature evolution of $\Delta$ and
$\Delta_{\rm pg}$ is the main goal of this work.
We show the results schematically in Fig.~\ref{fig:schematicE} and in more detail in
Figs.~\ref{fig:oneloopScale} and~\ref{fig:Escales}.}
\label{fig:SpectralPD}
\end{figure*}

In simple words, a precursor behavior to the SDW means the following:
In the SDW ordered state the Fermi surface gets reconstructed due to
 doubling of the unit cell,
 and  a gap
$\Delta (T)$ opens up for ``hot" fermions, whose Fermi momenta
$\kv_{hs}$ (see Fig.~\ref{fig:SpectralPD}~a) are connected by the SDW wave-vector $\Qv \approx (\pi,\pi)$. The
spectral function $A_{\kv_{hs}} (\omega)$ for such fermions has two $\delta$-function peaks at $\omega +\delta \mu
\approx \pm \Delta$, where $\delta \mu = \mu - \mu_0$, and $\mu$ and $\mu_0$ are the actual chemical potential and
the one
 for free fermions. A precursor to
  a SDW is a state above $T_N$, in which
the spectral function is continuous and non-zero for all $\omega$, yet there are maxima (humps) at energies
$\omega + \delta \mu \approx \pm  \Dpg$, where over some range of $T > T_N$, $\Dpg$ is comparable to $\Delta (T=0)$
(see Fig.~\ref{fig:SpectralPD}).
A convention, widely used in the interpretation of photoemission
 results, is that
  pseudogap behavior holds when
  the spectral function of a hot
fermion
 has two peaks at a finite frequency, and a normal
metallic behavior holds when it has a single  peak  at zero frequency.

We emphasize that precursor behavior
 is different from a non-Fermi liquid behavior caused by  coupling
to soft overdamped spin fluctuations. The latter gives rise to strong frequency dependent self-energy, which
 distributes the spectral weight over a wide range of frequencies. Yet, the  maximum of $A_{\kv_{hs}} (\omega)$
   remains at $\omega=0$.

To see how both non-Fermi liquid and precursor behavior emerge within the spin-fluctuation scenario,
 consider  a hot fermion, whose
  energy
  $\epsilon_{\kv_{hs}} = \epsilon_{\kv_{hs}+\Qv} = \mu_0$,
   and analyze the one-loop self-energy due to spin
  fluctuation exchange~\cite{Abanov2003}.
On the Matsubara axis,
 $\Sigma (\kv_{hs},\omega_n) = \int \diff \Omega_m \diff \qv \, G(\kv_{hs}+\qv , \omega_n+\Omega_m) \chi (\qv,
 \Omega_m)$,
  up to a numerical factor,  where $G$ and $\chi$ are fermionic and spin-fluctuation propagators, respectively
 (we define $\Sigma$ via $G^{-1} = G^{-1}_0 - \Sigma$).

  In a SDW state,   $\chi (\qv, \Omega)$ contains the $\delta$-function piece $\Delta^2 \delta (\Omega_m) \delta(\qv -\Qv)$, and the
   self-energy
    is
    $\Sigma (\kv_{hs}, \omega_n) = \Delta^2 G( \kv_{hs} +\Qv, \omega_n) \approx
  \Delta^2/(i\omega_n - (\epsilon_{\kv_{hs}+\Qv} - \mu)) = \Delta^2/(i\omega_n +\delta \mu)$.
   On
    the real frequency axis, this self-energy has a pole at
     $\omega =-\delta \mu - i0$.
   Using
    $G^{-1} (\kv_{hs}, \omega) = \omega +i0 + \delta \mu - \Delta^2/(\omega+i0 + \delta \mu)$, one immediately finds
    that the spectral function $A_{\kv_{hs}} (\omega)$  has two peaks at
  $\omega + \delta \mu = \pm \Delta$.  A precursor to SDW
  in the paramagnetic state
  emerges when the
   self-energy  still has a
    pole at a finite  $\omega =- \delta \mu$,
     but the pole moves to the
   lower frequency half-plane due to  finite damping.

   At $T=0$
    this does not happen because
     dynamical spin fluctuations
     are Landau
   overdamped and are slow modes compared to fermions. In this situation, the leading term in the
   self-energy is the convolution of the local Green's function, integrated over the momentum component perpendicular
   to the Fermi surface, and local bosonic propagator, integrated over the momentum that connects two points
   on the Fermi surface.
   This self-energy
     $\Sigma (\kv_{hs}, \omega_n) = \int \diff \Omega \, G_L (\omega_n + \Omega_m) \chi_L (\Omega_m)$ strongly
     depends on frequency and gives rise to  a
     redistribution of the
     spectral weight away from $\omega =0$, but it has no pole.

The situation changes at a finite $T$. Now integration over $\Omega_m$ is replaced by  summation over $\Omega_m =
2\pi m T$,
 and the self-energy contains the thermal contribution
 from  static SDW fluctuations.  The corresponding self-energy is $\Sigma_{\rm th} =
 T  \int \diff \qv \,G( \kv_{hs} + \qv, \omega) \chi (\qv, 0)$.
   It is natural to assume  that near a SDW instability, $\chi (\qv, 0)$  has an Ornstein-Zernike form $\chi (\qv,0)
   \propto \left[ (\qv-\Qv)^2 + \xi^{-2} \right]^{-1}$, where $\xi$ is the magnetic correlation length. The integral
   $\int \diff \qv\, \chi (\qv,0)$ is then confined to small $\qv-\Qv$ in dimensions $d \leq 2$.
 To first approximation one can then replace $G( \kv_{hs} + \qv, \omega)$ by $G( \kv_{hs} + \Qv, \omega)$ and move it
 out of   momentum    integral.
   One then   obtains the same $\Sigma_{\rm th} =   \left(\tDpg\right)^2/(\omega + i0 + \delta \mu)$ as in the SDW state, with   $\left(\tDpg\right)^2 =   T \int \diff \qv \, \chi (\qv,0)$.
   This form is indeed an approximate one as replacing $G( \kv_{hs} + \qv, \omega)$ by $G( \kv_{hs} + \Qv, \omega)$
   and moving it out from the momentum integral is only approximately correct when $\chi (\qv,0)$ is not a
   $\delta$-function.
    In more accurate one-loop calculations~\cite{Vilk1996,Randeria1998,Chubukov2007} the pole in $\Sigma (\kv_{hs},
    \omega)$ moves to the lower half-plane or transforms into a branch cut of the complex frequency.
This gives rise to broadening of the peak in the spectral function,  yet the maximum at $\omega +\delta \mu = \pm
\tDpg$,  survives in a finite $T$ range.
   For $d=2$, which we consider below, $\int \diff \qv \,\chi (\qv,0) \propto \log{\xi}$, and $\tDpg \sim \sqrt{T \log \xi}$.

Pseudogap behavior  at a finite $T$ in 2D has been extensively studied numerically in the last few
years~\cite{TremblayReview2006,Gunnarsson2015,Gull2015, WuWei2017,Schafer2021,
Simkovic2022}, using various modern computational
techniques for the Hubbard model, and was clearly detected at half-filling.
The fluctuation diagnostics method identified static
antiferromangetic fluctuation as the  source of the pseudogap behavior
~\cite{Gunnarsson2015}.
An identification of the pseudogap scale with the one-loop $\tDpg$ is
a more subtle issue.
$\Delta_{\rm pg}$, extracted from the numerical data,  depends only weakly on temperature in a
  finite temperature window above $T_N$  (Ref.~\cite{Schafer2021}),  while
  $\tDpg \sim T \log{\xi} $ contains $T$ as an overall scale.
     The authors of~\cite{Schafer2021} argued that their data for
   the magnetic correlation length are consistent with the exponential behavior $\xi \propto e^{T_0/T}$.
    Then     $\log{\xi} \sim 1/T$ compensates the overall $T$, and     $\tDpg$ becomes $T$-independent, like the measured $\Delta_{\rm pg}$.
    However, the exponential temperature dependence of $\xi$
      holds in a 2D
        Heisenberg model for localized spins~\cite{SachdevBook},
         but there is no obvious reason why it should hold in a metal.
    Indeed, using
     the one-loop approximation for the
     spin susceptibility, one obtains that $\xi$ only weakly depends on $T$, hence $\tDpg$  scales
     roughly as $T$, in  disagreement with the numerical data.

Another issue is the location of the pseudogap phase.  At a first glance, it should exist at a finite $T$ as long as spin correlation length is large, even if the ground state is not magnetically ordered. However, extensive quantum Monte-Carlo studies of fermion-boson models with a paramagnetic ground state found no evidence for the pseudogap~\cite{BergMC2, BergMC3}.  Recent numerical studies  of the Hubbard model at a finite doping also argued that pseudogap phase at a finite $T$ exists only in the range of  dopings where the ground state possesses some magnetic order~\cite{Simkovic2022}.

The goal of this work is to resolve these issues. For this we adopt the computational technique known
as the eikonal approach,  which allows one to sum up thermal contributions to the fermionic Green's function up to an infinite order.
To the best of our knowledge, the eikonal approach has been first applied in the solid state context  in the study of one-dimensional (1D) systems with charge density wave (CDW) fluctuations~\cite{Sadovskii1974,*Sadovskii1974b,*Sadovskii1979,*[{For a detailed discussion of the formalism, see }]SadovskiiBook} (see also Refs.~\cite{McKenzie1996}).
In the context of SDW fluctuations, the technique has been applied to analyze how the pseudogap survives when long-range magnetic order  gets destroyed
  by thermal
 fluctuations~\cite{Sedrakyan2010,Ye2019pg}, and how
 thermal fluctuations lead to pseudogap formation when one departs from
a metal~\cite{Schmalian1998,Sadovskii1999,Schmalian1999,
*Sadovskii_extra,*Sadovskii_extra_1,Sadovskii_review}.
These last studies, however, used the magnetic correlation length $\xi$ as an input parameter.
Below we extend the eikonal approach to spin polarization in the paramagnetic phase, from which we extract the temperature dependent correlation length $\xi (T)$.
We show that the pseudogap behavior does develop above $T_N$, and the  pseudogap scale $\Delta_{\rm pg}$, extracted from the full Green's function, is comparable to the $\tDpg \propto (T \log \xi)^{1/2}$, where $\xi = \xi (T)$ is the fully dressed correlation length.  In a sizable range of $T$ above $T_N$, this $\xi(T)$ is, to a good accuracy, exponential in $1/T$, such that $\Delta_{\rm pg}$ is nearly independent on $T$.  This is consistent with  Ref.~\cite{Schafer2021}.
 We further show that when the ground state is non magnetically ordered, the pseudogap does not develop
  due to non-exponential, but still strong temperature variation of the full $\xi (T)$, which keeps the system in a weak coupling regime.

\subsection{Summary of the results}
\label{sec:summaryresults}

\begin{figure}
\includegraphics[width=1\columnwidth]{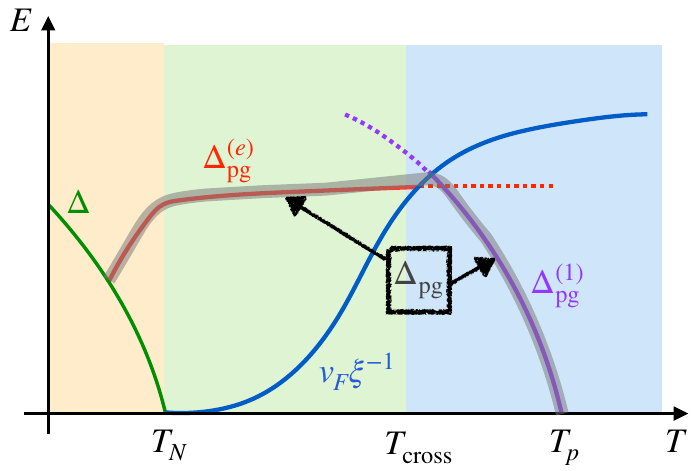}
\caption{
Energy scales that determine the evolution of the spectral function
with temperature:
the SDW order parameter $\Delta$ (green line), the pseudogap energy from the eikonal series $\Dpg^{(e)}$ (red line), the characteristic fermion energy $v_F \xi^{-1}$ (blue line) and the pseudogap energy from the one-loop calculation $\Delta^{(1)}_{\rm pg}$ (purple line). The energy evolution of the pseudogap scale $\Dpg$ over the whole temperature range
is highlighted in grey. The pseudogap emerges below $T_N$, remains almost a constant up to $T_{\rm cross}$, and then decreases and eventually vanishes at $T_p$. The regions below and above $T_{\rm cross}$ are termed ``strong" and ``weak" pseudogap regimes. Note that quantum fluctuations become relevant at roughly the same $T_p$, where the pseudogap disappears (see Sec.~\ref{sec:summaryresults} for the detailed discussion).
We show more specific results in Fig.~\ref{fig:oneloopScale} for $\Delta^{(1)}_{\rm pg}$ and  in Fig.~\ref{fig:Escales} for $\Delta, \Dpg^{(e)}$ and $v_F \xi^{-1}$.
\label{fig:schematicE}}
\end{figure}

We study thermal evolution of the spectral function
 in  the Hubbard model with
hopping $t$ between nearest and $t'$ between next-nearest neighbors,
 by varying $T$ and $U$ at a given hole doping $x > 0$.
  At large $U$, the relevant energy scale for
magnetic fluctuations is $J = 4t^2/U$.
 Like we said, we focus on ``hot" fermions, for which
 $\epsilon_{\kv} \approx \epsilon_{\kv + \Qv} \approx \mu_0$.

 Thermal fluctuations in the magnetically-ordered state at $T < T_N$ have been analyzed before, and we use these
 earlier results as input for our studies~\cite{Sedrakyan2010,Ye2019pg}. The strength of thermal fluctuations
  is controlled by the dimensionless parameter $\mathbb{t}^* = \frac{T}{J} |\log \epsilon|\propto \frac{T}{T_N}$,
  where $\epsilon$ is a deviation from  two-dimensionality (the parameter that cuts 2D logarithms at infinite $\xi$).
   Deep in the ordered phase at $T \ll T_N$, the spectral function of a fermion at a hot spot nearly vanishes below
   the scale set by the true SDW order $\Delta (T)$, and is peaked at $\omega + \delta \mu=\pm \Delta (T)$.
  (Fig.~\ref{fig:SpectralPD}(b)). In this regime, $\delta \mu$ is negative and   is comparable
   by magnitude
    to     $\Delta (T)$.
    For such low $T$, the one-loop mean field approximation works well.
   As $T$ increases, the SDW order parameter $\Delta (T)$ shrinks, and the spectral function displays two features:
    (i) a  true gap below $\Delta (T)$ (up to $e^{-\Delta (T)/T}$ corrections),
     and (ii) a hump at $\omega + \delta \mu = \pm \Dpg (T)>\Delta(T)$, where
    $\Dpg (T) \sim U\sqrt{\mathbb{t}^*} \sim U$ near $T_N$.
    The chemical potential
     is located
     between the SDW gap $\Delta (T)$ and the hump
      energy
       $\Dpg (T)$  (see Fig.~\ref{fig:SpectralPD}(c)).
  In the extreme case of $U$ much larger than the bandwidth, $\Dpg \approx U/2$ and $\delta \mu   \approx \mu   \approx -U/2$,   with corrections of order $J$.  The
  humps are then located at $\omega \approx U$ and at $\omega \sim -J$.
At $T = T_N$, $\Delta (T)$ vanishes and the spectral function becomes non-zero at all finite frequencies. Yet, the spectral function still has peaks at $\omega +\delta \mu = \pm \Dpg$.

The key result of our analysis is the identification of the
 system behavior in the paramagnetic phase.  We argue that the strength of the thermal contribution to the self-energy is determined by the dimensionless coupling
 $\lambda =\lambda (T) \propto T \xi^2 (T)$. A pseudogap behavior develops when $\lambda$ is larger than critical $\lambda_c = O(1)$.  This definitely holds above $T_N$, where $\xi (T)$ diverges.

  We argue that the proper description of thermal fluctuations at large $\lambda$
  requires one to sum up infinite series of diagrams for
 the  fermionic
   self-energy {\it and} for the polarization bubble, from which we extract the fully dressed correlation length
    $\xi (T)$.   The series can be viewed perturbatively
    as an expansion in
   $\mathbb{t}_0\sim \frac{T}{J}|\log \xi_0 |$, where $\xi_0$ is the    bare    magnetic correlation length.     In our calculations, we re-express the series in terms of
   $\mathbb{t}\sim \frac{T}{J}|\log \xi|$, where $\xi$ is the actual, fully renormalized correlation length, which we
   compute self-consistently.
   We explicitly sum up the series by converting them into certain integrals, which we evaluate analytically and obtain exact analytical formulas for the fully dressed fermionic Green's function and the correlation length.
   We find that the dressed $\xi$ is exponential in $T_0/T$, where
   $T_0\sim J$. The parameter $\mathbb{t} \sim (T/J) \log{\xi}$  is then $O(1)$,    which in turn justifies the need      to sum up infinite
    series of thermal     contributions to the self-energy and the polarization bubble.
  The fully dressed pseudogap scale, defined as $\Dpg^{(e)}$, scales as $(T \log \xi)^{1/2}$, like the one-loop pseudogap, and is almost independent on $T$.
   Its magnitude is the same as $\Dpg^{(e)}$ in the SDW state near $T_N$.
   These results are in agreement with the numerical data~\cite{Schafer2021}.
   We show the spectral function in this regime in Fig.~\ref{fig:SpectralPD}(d).
    It was termed a
   ``strong pseudogap regime", based on the analysis of the experimental data from various
   probes~\cite{NormalPseudogapReivew2005,Schmalian1998}

 For smaller $\lambda$, but still larger than the critical one,
 the self-energy
 due to thermal fluctuations changes because  one  cannot  pull fermionic Green's function out of the momentum integral.
  This in turn changes the behavior of the correlation length, which is no longer exponential in $1/T$.
   We argue that $\lambda (T)$ decreases with increasing $T$, and the pseudogap energy also decreases and
    eventually vanishes at $T = T_p$ (Fig. \ref{fig:schematicE}).  This regime was termed a ``weak pseudogap
       regime"~\cite{NormalPseudogapReivew2005,Schmalian1998}.  The shrinking and eventual vanishing of $\Dpg$
       is adequately described within the one-loop approximation.  We also argue that
        quantum spin fluctuations (the ones with non-zero bosonic Matsubara frequencies) become comparable to thermal ones starting from $T_q \sim T_p$, i.e.,  to a reasonable approximation the end point of the pseudogap behavior is also the boundary between thermal and quantum regimes.
        At $T > T_p$, the system displays a conventional metallic behavior, Fig.
     ~\ref{fig:SpectralPD}(e).

\begin{figure*}[t]
\subfigure[]{\includegraphics[width=0.66\columnwidth]{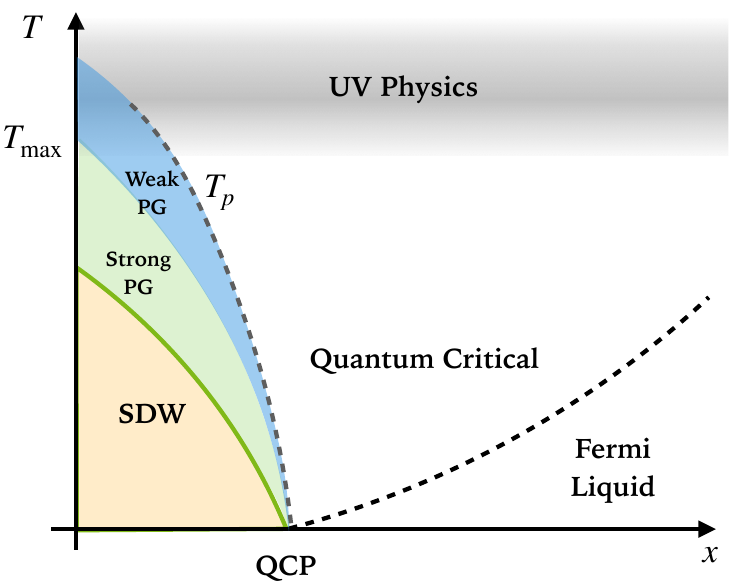}}
\subfigure[]{\includegraphics[width=0.66\columnwidth]{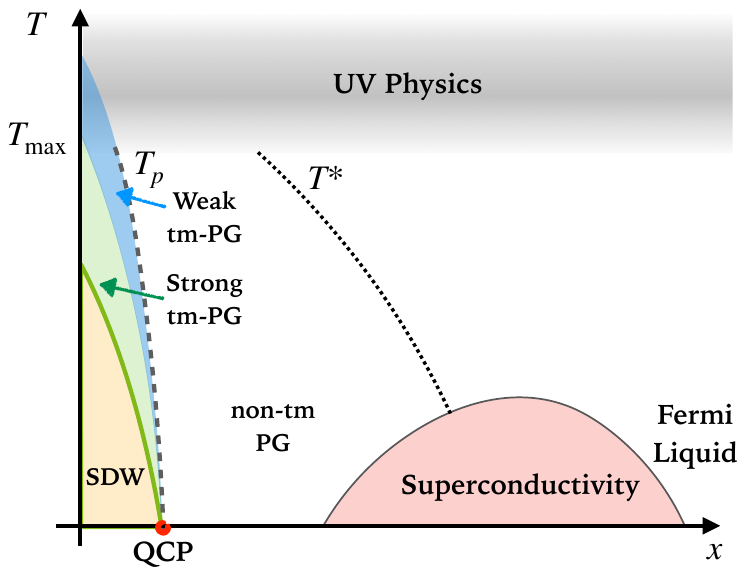}}
\subfigure[]{\includegraphics[width=0.66\columnwidth]{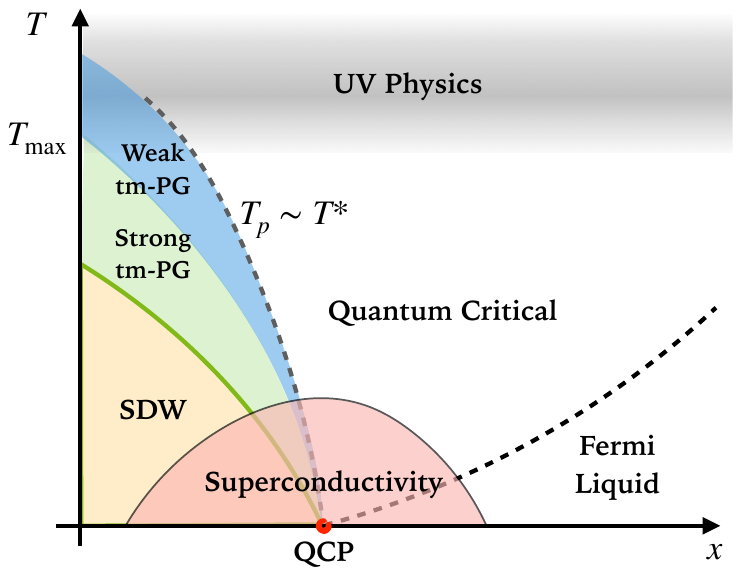}}
\caption{(a) A schematic phase diagram,
 obtained from
 our calculations (see Sec.~\ref{sec:phase_diagram}).
  A QCP corresponds to
  $x = x_c$
  ($U = U_c (x)$). There is a SDW order at smaller $x$ (larger $U$) and paramagnetic behavior holds for larger $x$ (smaller $U$).
  Pseudogap develops in the  regime, where the thermal contribution to the self-energy (the one from zero bosonic Matsubara frequency) is larger than the quantum one.  The temperature $T_p$, where pseudogap disappears, roughly coincides with the boundary of the thermal region.  In the quantum-critical region, thermal and quantum contributions to the self-energy are comparable in strength.  In the Fermi liquid region, Im $\Sigma (\omega) \propto \omega^2$.
(b,c): Two possible phase diagrams for the cuprates, based on our calculations. In both diagrams, the pseudogap behavior due to thermal magnetic fluctuations, denoted as ``tm-PG", develops only above  the $(\pi,\pi)$ ordered state.  In the strong pseudogap regime due to thermal fluctuations
 (Strong tm-PG), the pseudogap energy weakly depends on temperature, while in the weak thermal pseudogap regime
  (Weak tm-PG), it decreases with increasing $T$ and vanishes at $T^*$.  In (b), SDW order holds only at $x$ far smaller than the one for an optimal doping. This mimics the case of hole-doped curates.  We conjecture that the pseudogap behavior, observed  in hole-doped cuprates below  $T^* (x)$, which extends to near-optimal doping,
 is not caused by thermal spin fluctuations and is either a precursor to superconductivity or to Mott physics, or a different state of matter, possibly  with a topological order.  We label this regime as non-tm-PG.
In (c),
SDW order extends to near-optimal doping, and the boundary of the magnetic pseudogap, $T_p$, becomes close $T^*$.
 In this situation, the experimentally detected pseudogap behavior well may be due to thermal spin fluctuations.
This, we believe,  mimics the case of electron-doped cuprates. }
\label{fig:PhaseDiagramQT}
\end{figure*}

 Right above a QCP, we find that $\xi^{-2} (T)$ scales as $T$, modulo logarithms.  The coupling $\lambda$ is then
  independent on $T$. We find that its value is below the critical $\lambda_c$, hence pseudogap behavior does not emerge.
   The same holds  when $\xi$ is finite at $T=0$. Our results then show that pseudogap behavior emerges only when the ground state is magnetically ordered (Fig.~\ref{fig:PhaseDiagramQT}~a).
    This agrees with recent numerical study of the Hubbard model~\cite{Simkovic2022} and with quantum Monte Carlo analysis of a fermion-boson model near a $(\pi,\pi)$ SDW instability~\cite{BergMC1}.

We apply the results  to the cuprates and show the location of the pseudogap region due to thermal SDW fluctuations in Fig.~\ref{fig:PhaseDiagramQT}.  Most experiments indicate that in hole-doped cuprates a magnetic order is lost well
    before optimal doping.  Our results  indicate that in this situation, the observed pseudogap behavior below $T^* (x)$ in these materials is not due to  thermal magnetic fluctuations and is either the result of strong pairing fluctuations~\cite{Randeria1998,Millis1998,Berg2007,YMWu2021,Qi2022},
    or reflects a hidden, possibly topological order below $T^*$~\cite{Varma1997,Varma1999,Sachdev2018,YHZhang2020a,YHZhang2020b,Sachdev2022pga,Sachdev2022pgb}
    (Fig.~\ref{fig:PhaseDiagramQT}~b). If, however, a magnetic order (not necessary a $(\pi,\pi)$ one)  survives up to optimal doping, pseudogap behavior due to thermal magnetic fluctuations
     extends over a much wider range, and $T_p$, up to which this order holds, may be close to $T^*$
      (Fig.~\ref{fig:PhaseDiagramQT}~c). This last behavior holds in electron-doped cuprates, where a SDW order extends almost up to optimal doping
     ~\cite{ArmitageRMP}.

 The paper is organized as follows. In Sec.~\ref{sec:model} we introduce the model and review the mean field solution
for the magnetically ordered state. Here we list the results for the dynamical magnetic susceptibility, the Goldstone
modes, and the magnon-fermion vertex function. In Sec.~\ref{sec:thermal} we discuss the procedure to study the
pseudogap behavior from \emph{static thermal fluctuations}. We first review the one-loop results both in the
SDW-ordered phase and in the paramagnetic phase, and then discuss the eikonal approach, again first in the SDW phase
and then in the paramagnetic phase, where we also discuss infinite series for the spin polarization bubble, from
which we extract the temperature dependence of the correlation length. In Sec.~\ref{sec:EikonalResults} we present
our numerical solutions of eikonal equations for $\Delta (T)$, $\Dpg^{(e)} (T)$, and $v_F \xi^{-1}$ and analyze the
evolution of the fermion spectral function.
In Sec.~\ref{sec:phase_diagram} we locate the region, in which thermal fluctuations dominate,  on
 the phase diagram of the spin-fermion model on the $(T, x)$ plane, and  compare our phase diagram with the  experimental one  for high-$T_c$ cuprates.
  In Sec.~\ref{sec:summary} we summarize our findings.

\section{The model}
\label{sec:model}
The point of departure for our analysis is the one band Hubbard model for spin 1/2 fermions on a square lattice with
nearest and next-nearest neighbor hopping $t, t'$,
\begin{align}
\mac{H}_{\rm Hubbard}
 = \sum_\kv \sum_\sigma \epsilon_\kv a^{\dg}_{\kv, \sigma} a_{\kv , \sigma}
  +U\sum_i n_{i,\uparrow} n_{i,\downarrow},
\end{align}
where $\epsilon_\kv = -2t (\cos k_x + \cos k_y ) - 4 t' \cos k_x \cos k_y$.  For numerical calculations we set $t
=0.3 eV$ and $t' = -0.06 eV$.

 At small enough $U$, the ground state of $\mac{H}_{\rm Hubbard}$ is a Fermi liquid with a Fermi surface whose size
 is related to electron density $1-x$ by Luttinger theorem. We show the Fermi surface of non-interacting fermions in
 Fig.~\ref{fig:SpectralPD} (a). Near half-filling (at small $x$), the Fermi surface contains 8 special points called
 hot spots, for which $\kv$ and
  $\kv + \Qv = \kv + (\pi,\pi)$, are both on the Fermi surface ($\epsilon_\kv =
   \epsilon_{\kv+\Qv} = \mu_0$).
    For free fermions, the spectral function $A_\kv (\omega)$ at a hot spot is a $\delta$-function $\delta (\omega)$
   At finite $T$ and  $U$, the $\delta-$function broadens due to the fermionic self-energy, but remains peaked at
    $\omega =0$ (panel (e) in Fig. \ref{fig:SpectralPD}).

 We assume that at larger $U > U_c$,  the ground state at
 half-filling
 is a SDW state with ordering wave vector $\Qv$.
The value of $U_c$  is determined
    by
    solving
    the mean field equation for the SDW order parameter (See Fig.~\ref{fig:Ucx} for the solution of $U_c$ at different dopings).
       We further assume that the parameters are such that
     a commensurate SDW order holds at a finite doping $x$, up to a critical $x_c (U)$. We do not consider here an
     incommensurate spin order at a finite $x$,
     and stripe configurations, which emerge when an incommensurate order  melts down~\cite{Metzner2015, Dombre1990,
     Schulz1990, Shraiman1992, ChubukovMusaelian1995}.
  We describe the ground state and the finite temperature state proximate to the SDW by studying the mean field
  Hamiltonian and low energy fluctuations on top of it. In the strong coupling limit $U/t\gg 1$, this corresponds to
  the renormalized classical regime of the non-linear sigma model
     ~\cite{Chubukov1994}.
     Our main interest here
     is to study the physics in the intermediate coupling regime, when the Hubbard $U$ and the bandwidth are
     comparable.

We first review the mean field Hamiltonian, the low energy magnon dispersion,
 and the magnon-fermion coupling~\cite{Chubukov1997}.

The mean field Hamiltonian reads
\begin{align}
&\mathcal{H}_{\rm MF} =\sump_\kv
\sum_\sigma\\
&
\begin{pmatrix}
a^{\dg}_{\kv,\sigma} & a^{\dg}_{\kv+\ve{Q},\sigma}
\end{pmatrix}
\begin{pmatrix}
\e_{\kv} & -\Delta_0 \sgn{\sigma}\\
-\Delta_0 \sgn{\sigma} & \e_{\kv+\ve{Q}}
\end{pmatrix}
\begin{pmatrix}
a_{\kv,\sigma} \\
a_{\kv+\ve{Q},\sigma}
\end{pmatrix}\non
\end{align}
where $\Delta_0=\frac{U}{2} \langle \sum_\kv a^{\dg}_{\kv+\ve{Q}} \sigma_z a_{\kv} \rangle$ is the SDW order
parameter, $\sum_\kv$ and $\sum'_\kv$ denote the summation over the full and folded Brillouin zone, respectively, see
Fig. \ref{fig:SpectralPD}a.

The standard Bogoliubov transformation diagonalizes the mean field Hamiltonian
to
\begin{equation}
\mathcal{H}_{\rm MF} =\sump_\kv \e^v_\kv \gamma^{v\,\dagger}_{\kv,\sigma} \gamma^{v}_{\kv,\sigma} + \e^c_\kv
\gamma^{c\,\dagger}_{\kv,\sigma} \gamma^{c}_{\kv,\sigma},
\label{eq:meanfield0}
\end{equation}
where $\epsilon^{c,v}_\kv=\varepsilon^{+}_{\kv}\pm E_{\kv}$ with $\varepsilon^{+}_\kv =
\frac{\epsilon_\kv+\epsilon_{\kv+\Qv}}{2},\, \varepsilon^{-}_\kv = \frac{\epsilon_\kv-\epsilon_{\kv+\Qv}}{2}$ and
$E_\kv=\sqrt{\Delta_0^2+\left(\varepsilon^{-}_\kv\right) {}^2}$.
 The valence and conduction band operators
 $\gamma^v_{\kv,\sigma}$ and $\gamma^c_{\kv,\sigma}$
 are related to the original $a_{\kv,\sigma}$ and $
a_{\kv+\ve{Q},\sigma}$ as
\begin{align}
\begin{pmatrix}
a_{\kv,\sigma} \\
a_{\kv+\ve{Q},\sigma}
\end{pmatrix} &=
\mathsf{V}_{\kv,\sigma}
\begin{pmatrix}
\gamma^v_{\kv,\sigma} \\
\gamma^c_{\kv,\sigma}
\end{pmatrix},\non\\
\mathsf{V}_{\kv,\sigma} &=\begin{pmatrix}
\mathsf{v}_\kv & -\sgn(\sigma) \mathsf{u}_\kv\\
\sgn(\sigma) \mathsf{u}_\kv & \mathsf{v}_\kv
\end{pmatrix},
\label{eq:BogoliukovTransform}
\end{align}
$\mathsf{v}_\kv = \sqrt{\frac{1}{2} (1-\frac{\varepsilon^{-}_\kv}{E_\kv})}, \,  \mathsf{u}_\kv= \sqrt{\frac{1}{2}
(1+\frac{\varepsilon^{-}_\kv}{E_\kv})}$.

The low energy fluctuations in the SDW state are the Goldstone modes.
 They can be
   obtained by
   computing the magnetic susceptibility~\cite{Schrieffer1989,ChubukovFrenkel1992}.
    The propagators
      of
      the magnon mode $e_\qv$
       are
\begin{align}
\mathcal{D}^{+,(0)} (\qv, \Omega_m) &=- \langle T_\tau e_\qv (\tau) {e^{\dg}}_{\qv}(0) \rangle_{\Omega_m} =
\frac{1}{\iu \Omega_m-\Omega_\qv },\non\\
\mathcal{D}^{-,(0)} (\qv, \Omega_m) &=- \langle T_\tau {e^{\dg}}_\qv (\tau) e_{\qv}(0) \rangle_{\Omega_m} =
\frac{-1}{\iu \Omega_m+\Omega_\qv }
\label{eq:magnon0}
\end{align}
where $\langle \mathcal{O} (\tau) \rangle_{\Omega_m} = \int_0^{\beta} \diff \tau \eu^{\iu \Omega_m \tau} \langle
\mathcal{O}(\tau) \rangle$.
  In the small $t'/t$ limit, $\Omega_\qv \approx 4 J S \sqrt{1-\gamma_\qv^2}$,
  where  $ \gamma_\qv= \frac{1}{2} \left( \cos q_x + \cos q_y \right) $ and $J=4 t^2/U$.
    The dispersion
   $\Omega_\qv$ is gapless at $\qv= (0,0 )$ and $\qv=(\pi, \pi)$, corresponding to the two Goldstone modes
   of fluctuations
   transverse to the SDW order.

The electron-magnon coupling is
\begin{widetext}
\begin{align}
\mathcal{H}_{\rm el-mag}=\frac{U}{\sqrt{N}} \sum_{\kv,\qv} \left[\eta_{\qv} \left(e^{\dg}_{-\qv}+e_{\qv}\right)
a^{\dg}_{\kv+\qv,\sigma} a_{\kv,\sigma'}+\bar{\eta}_{\qv}
\left(e^{\dg}_{-\qv}-e_{\qv}\right)
{a^{\dg}}_{\kv+\qv,\sigma} a_{\kv+\bm{Q},\sigma'}\sgn(\sigma)\right]\delta_{\sigma,-\sigma'},
\label{eq:magfermion0}
\end{align}
\end{widetext}
where $\delta_{\sigma,-\sigma'}$ is present because magnons are transverse fluctuations (we set the SDW staggered
magnetization along ${\bf z}$).  The coherence factors are \ $\eta_\qv = \frac{1}{\sqrt{2}}
\left(\frac{1-\gamma_\qv}{1+\gamma_\qv}\right)^{1/4}, \bar{\eta}_\qv = \frac{1}{\sqrt{2}}
\left(\frac{1+\gamma_\qv}{1-\gamma_\qv}\right)^{1/4}$.
 We see that
the magnon-fermion coupling
scales as $\sqrt{|{\bf q}|}$ at small ${\bf q}$ and diverges as $1/\sqrt{|{\bf q}-{\bf Q}|}$ at ${\bf q}$ near ${\bf
Q}$.
 In terms of the
 conduction and valence fermions $\gamma^{c,v}$, this interaction is
\begin{widetext}
\begin{align}
\mathcal{H}_
{\rm el-mag}
=
\frac{U}{\sqrt{N}} \sump_\kv \sum_\qv \left \{\gamma^{c \dg}_{\kv \sigma} \gamma^v_{\kv+\qv,\sigma'}
\left[(\eta_{\qv} -\bar{\eta}_\qv )e^{\dg}_{\qv} + (\eta_{\qv} +\bar{\eta}_\qv )e_{-\qv}  \right] + h.c. \right
\}\delta_{\sigma,-\sigma'} +
...
\label{eq:magfermion1}
\end{align}
\end{widetext}
 where dots stand for the terms that involve
only conduction or only valence fermions.  Because the corresponding interaction vertices  are small in $|\qv|$, we
will not include these terms in
our analysis.

\begin{figure}
\includegraphics[width=0.8\columnwidth]{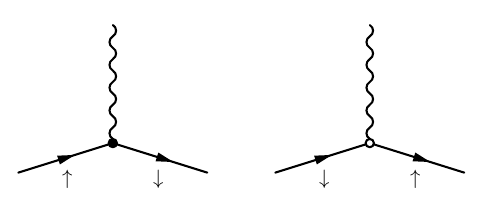}
\caption{Magnon-fermion vertex. The wavy line is used for magnon propagator, solid straight line for fermion
propagator, a filled (empty) circle $\bullet$ ($\circ$) for magnon-fermion vertex with outgoing spin-down (spin-up)
fermion and incoming spin-up (spin-down) fermion. Note that in the SDW state, each vertex must connect one conduction
fermion and one valence fermion.
\label{fig:MagnonFermion}}
\end{figure}

This interaction is illustrated graphically in Fig.~\ref{fig:MagnonFermion}. We use a wavy line for magnon
propagator, solid straight line for fermion propagator,
 a filled
  (empty) circle $\bullet$ ($\circ$) for magnon-fermion vertex with outgoing spin-down (spin-up) fermion and incoming
  spin-up (spin-down) fermion.

\section{Pseudogap from quasistatic spin fluctuations}
\label{sec:thermal}
In this section, we discuss
  how ``hot" fermions develop
  pseudogap behavior at a finite $T$ in both  SDW state and paramagnetic state,
  due to the singular self-energy contribution from thermal (static) spin fluctuations.
    We will identify a framework to study the effects of thermal fluctuations to infinite order in perturbation
    theory.
To set the stage for our analysis, in  Sec.~\ref{sec:oneloop} we first review and extend the one-loop calculation of the fermion
self-energy from thermal fluctuations and rationalize the need to include higher loop contributions.
In Sec.~\ref{sec:summation}, we discuss the computational procedure that allows one to sum up infinite series of
 thermal contributions to the fermionic self-energy and the bosonic polarization.
 This will allow us to determine self-consistently the fermionic Green's function,
  the chemical potential,
  the SDW order parameter, and the spin correlation length in the paramagnetic phase.

\subsection{
One-loop
 analysis}
\label{sec:oneloop}
The effects of quasistatic spin fluctuations have been studied both in the SDW state~\cite{Sedrakyan2010,Ye2019pg}
and in the paramagnetic state~\cite{Vilk1996,Vilk1997,Moca2000,Roy_2008,Schafer2021,Schmalian1998,Schmalian1999,Sadovskii1999,Sadovskii_review}. Here, we review
 and extend
one-loop calculations
in both phases
 and rationalize the need to
 include higher-loop contributions.
\begin{figure}[ht]
\subfigure[]{\includegraphics[width=0.45\columnwidth]{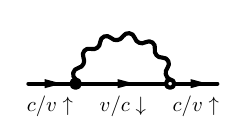}}
\subfigure[]{\includegraphics[width=0.45\columnwidth]{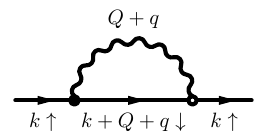}}
\caption{One loop self-energy. (a) Leading order diagram (at $\abs{\log \epsilon}$) in the SDW state. $c,v$ denote
the conduction and valence band fermions; $\uparrow, \downarrow$ denote the spin up and spin down state. (b) Leading
order diagram (at $\abs{\log \xi}$) in the paramagnetic state from magnetic fluctuations in the transverse channel.
\label{fig:OneLoop}}
\end{figure}

\subsubsection{SDW state}
For definiteness, consider the SDW ordered state at half-filling.
The thermal one-loop correction to the SDW order parameter
 diverges logarithmically in 2D and immediately destroys long-range SDW order, in agreement with Mermin-Wagner
 theorem~\cite{MerminWagner1966}.
 The one-loop fermionic self-energy is also logarithmically singular, but its effect is more nuanced, as we will see
 below. To circumvent the divergencies in 2D, we will consider the physics in dimension $2+\epsilon$ and use
 $\epsilon \ll 1$ to regularize the logarithmic singularity. Physically, systems with small but finite $\epsilon$ are
  highly anisotropic 3D systems with small hopping along the $z$ direction.

The one-loop correction to SDW order changes the order parameter $\Delta$ from $U/2$, which is its value at $T=0$ and
$U \gg t, t'$ to
\be
\Delta  = U\langle S_z \rangle = \frac{U}{2} \left(1-\bbt^*/2 \right)
\label{aa}
\ee
where
\beq
\bbt^* = \frac{2 T}{\pi J S} \ln \frac{\pi^2}{2\epsilon^2}
\label{zz_1}
\eeq
 is a dimensionless parameter, which measures the
strength of thermal fluctuations.

The one-loop self-energy for a conduction electron is given by the diagram in Fig.~\ref{fig:OneLoop}(a), using the
magnon-fermion coupling vertex in Eq.~\eqref{eq:magfermion1}.  In analytical form,
\begin{widetext}
\begin{align}
\Sigma^{c (1)} (\kv, \iu \omega_n) = & -U^2 \frac{T}{N} \sum_{\qv,m} G^{v (0)} (\kv+\qv, \iu \omega_n+\iu \Omega_m)
\left( (\eta_\qv - \bar{\eta}_\qv )^2 \mathcal{D}^{-,(0)} (\qv, \iu \Omega_m ) + (\eta_\qv + \bar{\eta}_\qv )^2
\mathcal{D}^{+,(0)} (-\qv, \iu \Omega_m ) \right)\non\\
\overset{\Omega_m = 0}{=} & - U^2 \frac{T}{N} \sum_{\qv } G^{v (0)} (\kv+\qv, \iu \omega_n) \left( (\eta_\qv -
\bar{\eta}_\qv )^2 \mathcal{D}^{-,(0)} (\qv,0 ) + (\eta_\qv + \bar{\eta}_\qv )^2 \mathcal{D}^{+,(0)} (-\qv,0 )
\right)  \non\\
= & \,\quad U^2 \frac{T}{N} \sump_{\qv }\frac{2(\bar{\eta}_\qv^2+\eta_\qv^2) }{\Omega_\qv}  G^{v (0)} (\kv+\qv, \iu
\omega_n) + \sump_{\qv+\Qv }\frac{2(\bar{\eta}_{\qv+\Qv}^2+\eta_{\qv+\Qv}^2) }{\Omega_{\qv+\Qv}}  G^{v (0)}
(\kv+\qv+\Qv, \iu \omega_n)  \non\\
 \approx &  \,\quad U^2 T \frac{2}{JS} \frac{|\ln \epsilon|} {2\pi}  G^{v (0)} (\kv, \iu \omega_n)
 \label{eq:oneloopSDW}
\end{align}
\end{widetext}
Here and below we define the sign of $\Sigma$ by requesting that $G^{-1} = G^{-1}_0 - \Sigma$.

In the second line of Eq.~\eqref{eq:oneloopSDW},
we kept only  the term with zero Matsubara frequency $\Omega_m =0$,  whereas in the last line we
present the
   result of the momentum integration with logarithmical accuracy, using
     $\int \diff^2 \qv\, \frac{1}{\Omega_\qv} \sim \int \diff^2 \qv \frac{1}{|\qv|} + \int \diff^2 \qv
     \frac{1}{|\qv-\Qv|} \sim |\log \epsilon|$ (a more accurate result is $\log{\pi/(\sqrt{2} |\epsilon|)}$).
The contribution to $\Sigma^{c(1)}$ from $G^{c(0)}$ is a subleading one, due to the gradient nature of the
electron-magnon coupling for small momentum transfer.
Substituting the form of $G^{v (0)} (\kv, \iu \omega_n)$ into (\ref{eq:oneloopSDW}), we find
\begin{align}
\Sigma^{c (1)} (\kv, \iu \omega_n) \approx \frac{\bbt^* (U/2)^2}{\iu \omega_n - (\epsilon^{v}_\kv - \mu)}.
\label{eq:Sigmac1}
\end{align}
A similar analysis for valence fermions yields
\begin{align}
\Sigma^{v (1)} (\kv, \iu \omega_n) \approx \frac{\bbt^* (U/2)^2}{\iu \omega_n - (\epsilon^{c}_\kv - \mu)}.
\label{eq:Sigmav1}
\end{align}
Treating $\Sigma$ perturbatively as a correction to the Green's function near its mass shell, we find that
at large $U$ each self-energy changes the fermionic energy from $\epsilon^{c,v} \approx \pm U/2 + \mathcal{O}(t)$ to
\be
\epsilon^{c,v} \approx \pm U \left( \langle S_z \rangle + \frac{\bbt^* }{4}\right)+\mathcal{O}(t)
\label{bb}
\ee
Substituting $ \langle S_z \rangle$ from (\ref{aa}), we find that the corrections of order $\bbt^*$ cancel out, hence
 the energies of the conduction and valence fermions remain $\epsilon^{c,v} \approx \pm U/2$.
 This feature has been interpreted as an indication that
  the gap between the conduction and the valence bands is the Hubbard $U$,
  set by Mott physics, and it survives even
  when $\langle S_z \rangle$ vanishes, despite the fact that at the mean-field level this gap is defined as
  $2 U \langle S_z \rangle$~\cite{Chubukov1997,Sedrakyan2010}.

\subsubsection{ Paramagnetic state}
\label{sec:oneloopdisorder}
Next, we consider the paramagnetic
 state at a finite temperature.
 To lowest order in $U$, the (Hartree-Fock) self-energy is purely static and renormalizes the hoppings and the
 chemical potential.   We move one step ahead and include into the self-energy multiple insertions of particle-hole
 bubbles.
  This effectively splits the interaction into charge and spin components.
   At the RPA level, the self-energy can be expressed as (Fig.~\ref{fig:OneLoop}(b))
      \begin{align}
   \Sigma (\kv, \omega_m) = -\frac{T}{2}  \sum_{m, \beta} & \int \frac{\diff^2 \qv}{(2\pi)^2} G (\kv + \qv, \omega_m
   + \Omega_m) \nonumber \\
   & \times \Gamma_{\alpha \beta;\beta\alpha} (\qv, \Omega_m)
   \label{cc}
   \end{align}
   where
      \begin{align}
   \Gamma_{\alpha \beta, \gamma \delta} (\qv, \Omega_m)  = \frac{U}{2} & \left(\frac{\delta_{\alpha\beta}
   \delta_{\gamma\delta}}{1 + U \Pi^{(c)} (\qv, \Omega_m)} \right. \nonumber \\
   & \left. - \frac{{\vec{ \sigma}}_{\alpha\beta} \cdot {\vec \sigma}_{\gamma\delta}}{1 - U \Pi^{(s)} (\qv,
   \Omega_m)}\right)
   \label{dd}
   \end{align}
   and $\Pi^{(c,s)} (q, \Omega_m)$ is the particle-hole bubble in the charge and spin channels~\footnote{This
   expression is obtained by collecting the
    renormalizations of the vertex function $\Gamma_{\alpha \beta, \gamma \delta} (K, P;P,K)$ ($K = ({\bf k},
    \omega_{m,k})$) that contain polarization bubbles $\Pi (K-P)$ and neglecting all other contributions.  This
    vertex
     function is different from  $\Gamma^\omega (K, P; K, P)$ which determines low-energy physics of a Fermi
     liquid.}. For example, without coupling with collective excitations, $\Pi^{(c,s)} (q, \Omega_m)$ can be
     determined from the convolution of two free fermion propagators, and satisfies $\Pi^{(c)} (q, \Omega_m)=
     \Pi^{(s)} (q, \Omega_m)=\Pi (q, \Omega_m)$.
     Near a SDW instability at $U \Pi^{(s)} (Q,0) =1$, the dominant interaction comes from spin fluctuations.
     Dropping the charge component of $\Gamma$, we obtain an effective model with the interaction mediated by spin
     fluctuations.
      Approximating the static $1- U\Pi^{(s)} (q)$ by Ornstein-Zernike form  $1- U\Pi^{(s)} (q) = \mathbb{c}
      \left(({\bf q} - {\bf Q})^2 + \xi^{-2}\right)$,  where $\mathbb{c}$ is a dimensionless constant, we obtain the
      thermal self-energy at a hot spot in the form~\cite{Vilk1996,Vilk1997,Moca2000,Roy_2008,Schafer2021}

    \begin{widetext}
\begin{align}
\Sigma^{(1)}_{\rm para}({\bf k}_{hs},i\omega_n)=
 & 3 \bar{g} T  \int \frac{\diff{\qv}}{(2\pi)^2}\frac{1}{\iu \omega_n -
 v_F \tilde{q}_{\perp}}\frac{1}{\tilde{q}_\perp^2 + \tilde{q}_\parallel^2 +\xi^{-2}} = -\iu \sgn \omega_m
 \frac{3\bar{g}T}{2\pi v_F \xi^{-1}} f(\frac{|\omega_n|
 }{v_F \xi^{-1}}).
\label{eq:oneloop0}
\end{align}
\end{widetext}
where $\bar{g} \sim \mathbb{c}^{-1} U$ is an effective coupling,
\beq
f(y)=\frac{1}{\sqrt{y^2-1}} \log \left(y+\sqrt{y^2-1} \right),
\eeq
 and the factor of $3$ comes from spin summation.
It is convenient to introduce the dimensionless coupling
\be
\lambda = \frac{3\bar{g}T}{2\pi (v_F \xi^{-1})^2}.
\label{zz_3}
\ee
and dimensionless frequency
$\mathsf{w}_m=\omega_m/(v_F \xi^{-1})$.
The Green's function is
\begin{align}
&G(\kv_{hs}, \omega_m) =\non\\
& \left(\iu v_F \xi^{-1}  \left(\mathsf{w}_m + \lambda \sgn \mathsf{w}_m \frac{\log \left(|\mathsf{w}_m|+\sqrt{(\mathsf{w}_m)^2-1} \right) }{{\sqrt{(\mathsf{w}_m)^2-1}}}  \right)\right)^{-1}.
\end{align}
To see the effect of the self-energy, it is instructive to convert this expression onto the real axis, $i\omega_m \to
\omega + i \delta$.
 The retarded self-energy is
\begin{align}
    &\Sigma_{\rm para}^{(1)}(\kv_{hs}, \omega) =\non\\
    & v_F \xi^{-1} \lambda \left( \frac{\log \left( \mathsf{w}+\sqrt{\mathsf{w}^2+1}\right)}{\sqrt{\mathsf{w}^2+1}}
    -\iu\frac{\pi/2}{\sqrt{\mathsf{w}^2+1}} \right),
    \label{eq:oneloopfull}
\end{align}
note that the real part of $\Sigma_{\rm para}^{(1)}(\kv_{hs}, \omega)$ is an odd function of $\mathsf{w}$.
The retarded Green's function is
\begin{align}
&G_{ret} (\kv_{hs},\mathsf{w}) = \non\\
&\left( v_F \xi^{-1} \left(\mathsf{w} - \lambda \frac{\log\left(\mathsf{w}+\sqrt{\mathsf{w}^2+1} \right)}{{\sqrt{\mathsf{w}^2+1}}}  + \iu \lambda \frac{\pi/2}{\sqrt{\mathsf{w}^2+1}}  \right)\right)^{-1}.
\label{ff}
\end{align}
\begin{figure}
\includegraphics[width=0.9\columnwidth]{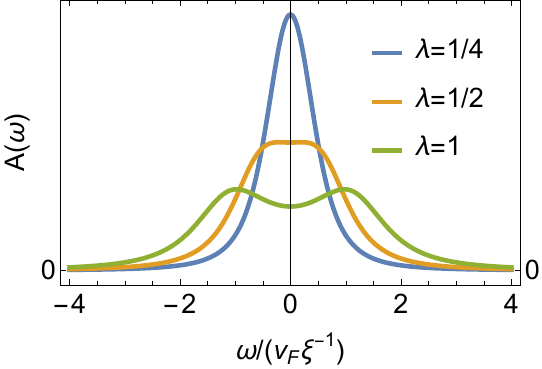}
\caption{Spectral function at the hot spot from the one-loop calculation (Eq.~\ref{ff}). As the dimensionless
coupling $\lambda = \frac{3\bar{g}T}{2\pi (v_F \xi^{-1})^2} $ increases, the spectral function shows pseudogap
behavior when $\lambda>\lambda_c=0.47$.
\label{fig:oneloop}}
\end{figure}

We plot
 the spectral function
  $A(\kv_{hs},\mathsf{w})=-(1/\pi)$ Im$G_{\text{ret}} (\kv_{hs},\mathsf{w})$ in Fig.~\ref{fig:oneloop}.
We see that at small $\lambda$, $A(\kv_{hs},\mathsf{w})$ is peaked at $\mathsf{w} =0$, as is expected for weakly
interacting fermions with momenta at the Fermi surface.
 The key effect of the self-energy at these $\lambda$ is to introduce a finite width of $A(\kv_{hs},\mathsf{w})$.
 However once $\lambda$ exceeds the critical value $\lambda_c \approx 0.47$, the maximum of $A (\kv_{hs},\mathsf{w})$
 shifts to a finite $|\mathsf{w}| \sim \sqrt{\lambda - \lambda_c}$, while at zero frequency,
 $A (\kv_{hs},0)$  now becomes a minimum (see Fig.~\ref{fig:oneloop}).
 This implies that thermal fluctuations in the paramagnetic state do give rise to pseudogap behavior
  already at one-loop order, if the coupling exceeds the threshold value. We define the one-loop pseudogap as
 $\tDpg$.

At small frequencies, the evolution of the spectral function with increasing $\lambda$ can be obtained analytically.
Expanding $A(\kv_{hs},\mathsf{w})$ near $\mathsf{w} =0$, we obtain
\beq
A(\kv_{hs},\mathsf{w}) = \frac{\lambda}{2 v_F \xi^{-1}} \frac{1}{\frac{\pi^2 \lambda^2}{4} +\mathsf{w}^2\left((1-\lambda)^2 -\frac{\lambda^2 \pi^2}{8}\right)}
\label{zz}
\eeq
We see that the maximum of $A(\kv_{hs},\mathsf{w})$ remains at $ \mathsf{w}=0$ as long as $\lambda(1 +\pi/(2 \sqrt{2}) <1$. The critical value $\lambda_c = 2\sqrt{2}/(2\sqrt{2} + \pi) = 0.4738$.

\begin{figure}
\includegraphics[width=0.9\columnwidth]{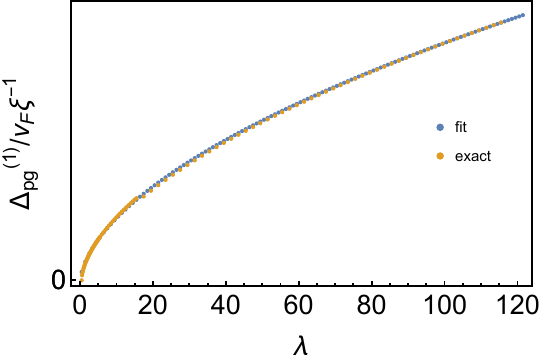}
\caption{
The ratio of the pseudogap scale $\tDpg$, extracted from the one-loop Green's function, and
$v_F \xi^{-1}$. Orange dots: the ratio, extracted from the numerical analysis of Eq.~\eqref{ff}.
Blue dots: the fit to the analytical expression
 $\Delta^{(1)}_{\rm pg}/(v_F \xi^{-1}) = \left(\frac{1}{2}\lambda \log \mathbb{b} \lambda\right)^{1/2}$.
   The best fit is for $\mathbb{b}= 14.8$.
\label{fig:oneloopScale}}
\end{figure}

  We next consider large $\lambda$. We assume and then verify that
   the position of the maximum of $A (\kv_{hs},\mathsf{w})$ moves to $\mathsf{w} \gg 1$,
 i.e., to $\omega \gg v_f \xi^{-1}$  For such $\mathsf{w}$, the momentum integration in the
   expression
   for the self-energy is fully confined to the bosonic
    term
    $\frac{1}{\tilde{q}_\perp^2 + \tilde{q}_\parallel^2 +\xi^{-2}}$, while the fermionic Green's function can be
    moved out of the momentum integral.
     For the retarded self-energy we obtain for such $\mathsf{w}$
      with logarithmic accuracy:
   \begin{equation}
\Sigma_{\rm para}^{(1)}(\kv_{hs}, \omega) = v_F \xi^{-1} \lambda \left( \frac{
\log{\mathsf{w}}}{\mathsf{w}} - \iu \frac{\pi}{2 \mathsf{w}}\right)
     \label{eq:oneloopfull_1}
\end{equation}
  The retarded Green's function is
  \begin{align}
&G_{ret} (\kv_{hs},\mathsf{w}) = \non\\
&\left( v_F \xi^{-1} \left(\mathsf{w} - \lambda \frac{\log\left(
\mathsf{w}\right)}{\mathsf{w}} +
 + \iu \lambda \frac{\pi}{2\mathsf{w}}  \right)\right)^{-1},
\label{ff_1}
\end{align}
 and the spectral function is
\beq
A(\kv_{hs},\mathsf{w}) = \frac{\lambda\mathsf{w} }{2 v_F \xi^{-1}} \frac{1}{\frac{\pi^2 \lambda^2}{4} +\left(\mathsf{w}^2 -\lambda \log{
\mathsf{w}} \right)^2}.
\label{zz_1a}
\eeq
This function has a maximum at $\mathsf{w} \approx \left(\frac{\lambda \log{
\lambda}}{2}\right)^{1/2}$.
  The corrections to this expression are of order
  $\log(\log(\lambda))$. They change the prefactor for $\lambda$ under the logarithm
   to $\log \left(\mathbb{b}(\lambda) \lambda\right) $, where $ \mathbb{b}(\lambda)$ is a slowly varying function of $\lambda$.
    In Fig.~\ref{fig:oneloopScale} we plot  $\tDpg/(v_F \xi^{-1})$, which we obtained numerically, without expanding at large $\mathsf{w}$, along with the analytical
     $\tDpg/(v_F \xi^{-1})  = \left(\lambda \log{( \mathbb{b}\lambda)}\right)^{1/2}$
    We found a good match  by setting $\mathbb{b}=14.8$ independent on $\lambda$.
We emphasize that $\tDpg/(v_F \xi^{-1})$  is large at large $\lambda$, which justifies the assumption that we used to obtain (\ref{zz_1}).

Because $\lambda \propto \xi^{-2}$,  the $(\lambda \log{\lambda})^{1/2}$ dependence at large $\lambda$ can be approximated by a more simple $\lambda \abs{\log {\xi}}$,  which is more convenient for calculations beyond one-loop order.
In  dimension $d = 2 + \epsilon$, $\log \xi$ is replaced by $\mathsf{L} = 0.5 \log \frac{\pi^2}{2(\epsilon^2 + \xi^{-2})} $. Using these modifications and extending the result to $\kv$ near, but not necessary at a hot spot,
 we obtain at large $\lambda$
 \be
 \Sigma_{\rm para}^{(1)}(\kv, \omega) \approx (v_F \xi^{-1})^2 \frac{\lambda \mathsf{L}}{
 \omega -\left(\epsilon_{\kv+\Qv} - \mu\right)}.
 \label{ii}
 \ee
 At $\xi^{-1}=0$, this expression has the same form
 as one-loop self-energy at the end point of the SDW state, Eqs. (\ref{eq:Sigmac1}) and (\ref{eq:Sigmav1}),  once we
 set $\Delta\rightarrow 0^+$ and identify
 $\epsilon_\kv^v \rightarrow \min\{ \epsilon_\kv, \epsilon_{\kv + \Qv} \}, \epsilon_\kv^c \rightarrow \max\{
 \epsilon_\kv, \epsilon_{\kv + \Qv} \}$.   The prefactors in (\ref{ii}) and  (\ref{eq:Sigmac1})-(\ref{eq:Sigmav1})
 match if we set  the prefactor $\mathbb{c}$ in
  $(1-U \Pi^{(s)} ({\bf q},0)) = \mathbb{c} \left(\xi^{-2} + ({\bf q} - {\bf Q})^2 \right)$   to be    $\mathbb{c} \sim J/U
  \sim (t/U)^2$.
    While $\mathbb{c}\sim (t/U)^2$ holds deep in the SDW state when $\Delta \sim U/2$, we will not discuss here how
    to justify the expression for $\mathbb{c}$ in the paramagnetic state and use it as a phenomenological element of
    our analysis.
   With the choice of $\mathbb{c}\sim J/U$ as the proper prefactor, the one-loop self-energy in the paramagnetic
   state
    at $|\omega - (\epsilon_{\kv+\Qv} -\mu)| \gg v_F \xi^{-1}$ is
   \begin{align}
\Sigma^{'(1)}_{\rm para}(\kv, \omega ) = \frac{ \bbt (U/2)^2 }{\omega - (\epsilon_{\kv + \Qv}-\mu)},
\label{eq:Sigma1}
\end{align}
where
\beq
\bbt = \frac{4 T}{\pi J S} L = \frac{2 T}{\pi J S} \ln \frac{\pi^2
}{2(\epsilon^2 + \xi^{-2})}.
\eeq
At $\xi^{-1} =0$, this coincides with $\bbt^*$ from Eq. (\ref{zz_1}).
The pseudogap energy $\tDpg = (U/2) \sqrt{\bbt}$.

The correlation length at the one-loop order is given by
  \be
  \xi = \left(\frac{A}{1-U \Pi^{(s)} ({\bf Q},0)}\right)^{1/2},
  \label{mm}
 \ee
where the polarization bubble $\Pi^{(s)} ({\bf Q},0)$ is constructed out of Green's functions of free fermions.
 Evaluating $\Pi^{(s)} ({\bf Q},0)$  for $t-t'$ dispersion, we find that $\xi$ is weakly temperature dependent.
 We label this correlation length as $\xi_0$ later in the text to distinguish it from the fully dressed $\xi$, which, as we will show, is strongly $T-$dependent.

\subsection{Rational to go beyond one-loop analysis}

 We now rationalize the need to
  go beyond the one-loop analysis

  For the SDW state, the one-loop formulas for $\langle S_z \rangle$ and for the self-energy  are the leading terms in an expansion in
$\bbt^*$. Meanwhile, the SDW order vanishes at $T=T_N$, at which $\bbt^* = O(1)$, as is clear from (\ref{aa}).  To understand
  how  $\langle S_z \rangle$  evolves at these $\bbt^*$ values, we clearly need to include terms
  beyond the one-loop order.
   The same holds for the fermionic self-energy, which also evolves at $\bbt^* = O(1)$.
   One can easily verify that for both $\langle S_z \rangle$  and the self-energy,
  $n-$loop order terms are of order
  $({\bbt^*})^n$, i.e., terms up to an infinite loop order
   have to be included in the analysis at $\bbt^* = O(1)$.

 The rational for  the paramagnetic phase is similar.
 First, at $\bbt = O(1)$,  higher-loop diagrams for the thermal self-energy are of the same order as the one-loop one, and have
 to be kept.  Second, the corrections to the polarization $\Pi^{(s)} ({\bf Q},0)$, which determines the correlation length
  via  (\ref{mm}), also scale as powers of $\bbt$ and should all be kept at $\bbt = O(1)$.
In this respect, there is a
similarity between the correlation
     length $\xi$ in the paramagnetic state and
     $\langle S_z \rangle$
     in the SDW state -- both have to be computed
      by summing up infinite series in either $\bbt$ or $\bbt^*$.

\subsection{Infinite summation in the quasistatic limit}
\label{sec:summation}
In this subsection, we collect contributions to order $(\bbt^*)^m$ and $\bbt^m$ with $m$ up to infinity
 and  apply the eikonal formalism  to obtain fully dressed variables.

    For the SDW state, the diagrammatic    series are determined kinematically by the structure of the electron-magnon
    coupling in Eqs.~\eqref{eq:magfermion0} and~\eqref{eq:magfermion1}.
We show below that the analysis of the full spectral function
 shows that there are two energy scales.
  One is the fully renormalized SDW order parameter $\Delta = U \langle S_z \rangle$,
  below which $A(k,\omega)$ vanishes,  and the other is the pseudogap scale
  $\Dpg^{(e)}$, where the spectral function has a hump.

   In the paramagnetic state, there are again two energy scales, the pseudogap $\Dpg^{(e)}$
 in the full Green's functions, obtained in the eikonal approach,   and $v_F \xi^{-1}$, where $\xi$ is
   given by (\ref{mm}) with the fully renormalized polarization bubble.
    We show that $\Dpg^{(e)}$ is comparable to the one-loop $\tDpg$,  but the full $\xi = \xi (T)$ differs from one-loop result and is strongly $T-$dependent.\\

\subsubsection{SDW state}
\begin{figure*}[t]
\includegraphics[width=2\columnwidth]{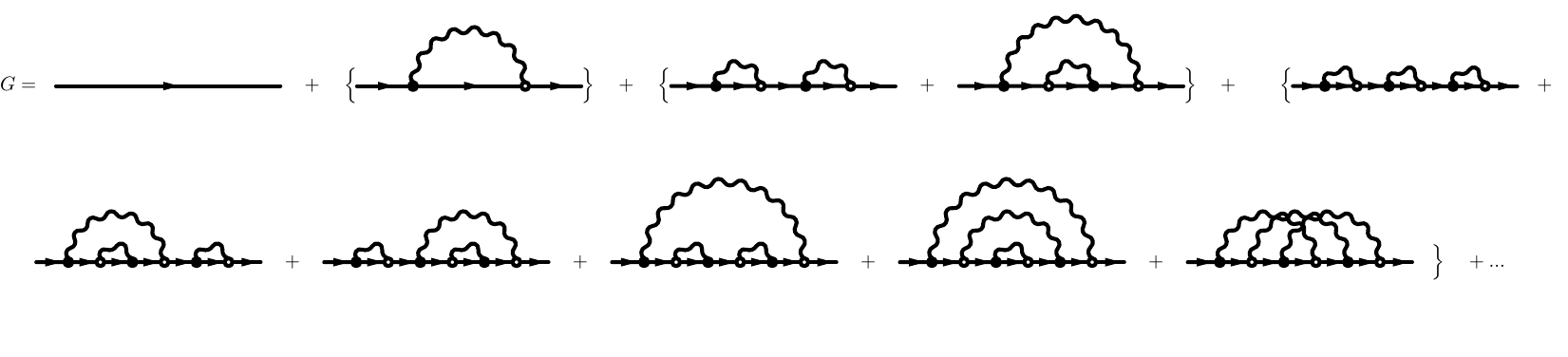}
\caption{Diagrammatic series for the full Green's function
 up to three loop order. In the SDW state,
 logarithmic corrections come from  the transverse magnetic fluctuations, but not from the longitudinal ones. Keeping only transverse fluctuations, we     obtain the combinatoric factor for the set of equivalent m-th order diagrams
      $\mathcal{C}_m=m!$ (see the discussion below Eq.~\eqref{jj}). In the paramagnetic state, magnetic fluctuations are $SU(2)$-symmetric, i.e., all three components of the magnetic susceptibility equally contribute to the self-energy.
       We verified, however, that the result for the self-energy does not change qualitatively if we only keep $n=2$ components. In this case, for which we report the results, $\mathcal{C}'_m = m!$ also in a paramagnetic state.}
\label{fig:Zinfty}
\end{figure*}

In the SDW state, the full Green's function $G^{c,v}$ with self-energy corrections
to all loop orders reads
\begin{widetext}
\begin{align}
G^{c,v}(\kv,\iu \omega_n) = G^{c,v (0)}(\kv, \iu \omega_n) \sum_m \mathcal{C}_m \left( \bbt^*
\frac{U^2}{4}\right)^{m}\times
 \left(G^{v,c (0)}(\kv, \iu \omega_n)  G^{c,v (0)}(\kv, \iu \omega_n) \right)^m.
\label{jj}
\end{align}
\end{widetext}
The combinatoric factor $\mathcal{C}_m = m!$ is determined by counting
the number of non-equivalent diagrams of order $(\bbt^*)^m$
at the $m-$th loop order.  The number is set by the  structure of the fermion-magnon vertices in
Eq.~\eqref{eq:magfermion1}, which  requires that
 each magnon propagator must be attached to one solid circle vertex ($\bullet$) and one empty circle vertex
 ($\circ$).
This requirement is due to the spin conservation, i.e.\ the $U(1)$ spin rotation symmetry in the collinear SDW state.
In Fig.~\ref{fig:Zinfty}, we show the diagrammatic series for the full Green's function up to
 three-loop
 order.

The series in Eq. (\ref{jj})
 can be summed exactly by using $\mathcal{C}_m = \Gamma(m+1)= \int_0^\infty \diff x \,x^{m}\eu^{-x}$ and expressing
  the full Green's function in the integral form as
  \begin{align}
G^{c,v}(\kv,\omega) = G^{c,v(0)}(\kv, \omega) \int_0^\infty \diff t \eu^{-t} \frac{1}{1-u_{\kv,\omega} t},
\label{eq:fullGreenSDW}
\end{align}
where $u_{\kv,\omega} = \bbt^* (U/2)^2 G^{c(0)}(\kv,\omega) G^{v(0)}(\kv,\omega)$
 and
 $G^{c,v(0)}$ are the bare Green's functions for conduction and valence fermions, but with the exact chemical
 potential $\mu = \mu(x,T)$ and fully renormalized SDW order parameter $\Delta = \Delta(x,T) = U \langle S_z\rangle$.
To determine the chemical potential
and the SDW order parameter,
we express the fermion density and $\Delta(x,T)$ in terms of  the fermionic spectral functions
$A^{c,v}_\kv(\omega)  = -\frac{1}{\pi} \im G^{c,v}(\kv, \omega+ \iu \delta)$ as~\cite{Sedrakyan2010}
\begin{align}
\frac{1-x}{2}  &=  \int \diff \omega \int \frac{\diff^2 \kv}{(2\pi)^2} n_F (\omega) A_\kv (\omega)\non\\
& =  \int \diff \omega \int \frac{\diff^2 \kv}{(2\pi)^2} n_F (\omega) \left(\mathsf{u}_\kv^2  A^c(\kv, \omega) +
\mathsf{v}_\kv^2  A^v(\kv, \omega)\right),\non\\
\langle S_z \rangle &= \int \diff \omega \int \frac{\diff^2 \kv}{(2\pi)^2} n_F(\omega) \mathsf{u}_\kv \mathsf{v}_\kv
\left(A_\kv^c(\omega)-A_\kv^v(\omega)\right),
\label{eq:selfconsistentSDW}
\end{align}
where
$n_F(\omega) = 1/(\exp(\omega/T)+1)$ is the Fermi function, and the coherence factors $\mathsf{u}_\kv,
\mathsf{v}_\kv$ are the same as in Eq.~\eqref{eq:BogoliukovTransform},  but with $\Delta_0$ in $E_\kv$ replaced by
$\Delta=\langle S_z \rangle /U$.

Solving Eq. \eqref{eq:fullGreenSDW} we find
\begin{widetext}
\begin{align}
G^{c,v}(\kv,\omega) = G^{c,v(0)}(\kv, \omega) \frac{e^{-1/u_{\kv,\omega}}}{u_{\kv,\omega}} \left({\rm{Ci}} \left(\frac{1}{u_{\kv,\omega}}\right) + {\rm{Si}} \left(\frac{1}{u_{\kv,\omega}}\right) - \iu \pi \sgn \left(\Im \left( \frac{1}{u_{\kv,\omega}}\right) \right)\right),
\end{align}
\end{widetext}
 where ${\rm{Ci}} (...)$ and ${\rm Si} (..)$ are CoshIntegral and SinhIntegral.  Substituting the expression for $u_{\kv,\omega}$ and evaluating the imaginary part of the full Green's function, we obtain
\beq
A(\kv,\omega)=\frac{\abs{\bw+\emi^-_\kv}}{\mathbb{t}^* (U/2)^2}\exp\left[-\frac{\bw^2-E_\kv^2}{\mathbb{t}^*
(U/2)^2}\right] \Theta(\bw^2-E_\kv^2)
\label{kk}
\ee
 where $\bw=\omega+\mu-\emi^+_\kv$, $E_\kv=\sqrt{(\emi^-_\kv)^2+\Delta^2}$, and we remind the reader that
 $\varepsilon^{+}_\kv = \frac{\epsilon_\kv+\epsilon_{\kv+\Qv}}{2}$ and $\varepsilon^{-}_\kv =
 \frac{\epsilon_\kv-\epsilon_{\kv+\Qv}}{2}$.
 At a hot spot, $\varepsilon^{-}_\kv = 0$, hence $E_k = \Delta$, and
 $\bw=\omega+\delta \mu$, where $\delta \mu = \mu - \mu_0$, the latter being the chemical potential of free fermions.
  The spectral function as a function of $\bw$ vanishes at $|\bw| <\Delta$, and has a maximum at
  $|\bw| = \Dpg^{(e)}$, where $\Dpg^{(e)} =\frac{1}{\sqrt{2}}(U/2) (\mathbb{t}^*)^{1/2}$. For $\Delta > \Dpg^{(e)}$, the spectral function shows two
  peaks at  $\pm \Delta \approx \pm \Delta_0$ (see Fig.~\ref{fig:SpectralPD} (b)), and when $\Delta < \Dpg^{(e)}$,
  the spectral function vanishes below the SDW scale $\Delta$ and displays the humps at $\pm \Dpg^{(e)}$, which is
  comparable to $\Delta_0$.
   (see Fig.~\ref{fig:SpectralPD} (c)).
In one further includes the non-thermal contribution to the fully renormalized fermion Green's function,
     finite jumps in the spectral function at $\bar{\omega} =\pm \Delta$ likely become the peaks~\cite{Sedrakyan2010}. Substituting the spectral function into~\eqref{eq:selfconsistentSDW}, we find the self-consistent equations for
 $\mu(x,T)$ and $\Delta(x,T)$ in an integral form as
\begin{widetext}
\begin{align}
\frac{1-x}{2}&=\int \diff \omega \int \frac{\diff \kv}{(2\pi)^2}
\frac{\abs{\bw+\emi^-_\kv}}{\mathbb{t}^* (U/2)^2}\exp\left[-\frac{\bw^2-E_\kv^2}{\mathbb{t}^* (U/2)^2}\right]
\Theta(\bw^2-E_\kv^2) n_F(\omega) \label{eq:selfconsistentSDW1}\\
\frac{1}{U}&=\int \diff \omega \int \frac{\diff \kv}{(2\pi)^2} \frac{-\sgn \bar{\omega}}{\mathbb{t}^* (U/2)^2}
\exp\left[-\frac{\bw^2-E_\kv^2}{\mathbb{t}^* (U/2)^2}\right]
\Theta(\bw^2-E_\kv^2)n_F(\omega)\label{eq:selfconsistentSDW2}
\end{align}
\end{widetext}
Analyzing the equations analytically, we find that the SDW gap at $T=0$ splits at finite $T$ into
  $\Dpg^{(e)}$,
  determined by the
     argument of the exponent
     in Eqs.~\eqref{eq:selfconsistentSDW1} and~\eqref{eq:selfconsistentDis1}, and $\Delta$, determined by the
     $\Theta$-function in Eq.~\eqref{eq:selfconsistentSDW1}.

We present the numerical results for $\Dpg^{(e)}$, $\Delta$, and the spectral function in Sec. \ref{sec:EikonalResults}.

\subsubsection{
 Paramagnetic state}
In the
paramagnetic state,  the full Green's function is expressed as
\begin{widetext}
\begin{align}
G(\kv,\iu \omega_n) = G^{(0)}(\kv, \iu \omega_n) \sum_m \mathcal{C}'_m \left( \bbt \frac{U^2}{4}\right)^{m}\times
 \left(G^{ (0)}(\kv, \iu \omega_n)  G^{ (0)}(\kv+\Qv, \iu \omega_n) \right)^m.
\label{eq:twopoint2}
\end{align}
\end{widetext}
The combinatoric factor is determined by the
 number $n$ of fluctuating spin components.
 For isotropic fluctuations, $n=3$, and $\mathcal{C}'_m=(2m+1)!!$~\cite{Schmalian1999,Sedrakyan2010};
  for transverse fluctuations, $n=2$, and $\mathcal{C}'_m=m!$, the same as in the SDW state.
  We verified that in both cases the system develops pseudogap behavior, the difference being only quantitative.
    Since the  $n=2$ case is simpler from computational perspective, below we present the results
    for $n=2$    (the coupling $\lambda$ for $n=2$ is the same as in (\ref{zz_3}), but with the overall factor $2$ instead of $3$). The Green's function  $G(\kv,\iu \omega_n)$ can be presented in an integral form, similar to
 Eq.~\eqref{eq:fullGreenSDW}:
\begin{align}
G(\kv,\omega) = G^{(0)}(\kv, \omega) \int_0^\infty \diff t \eu^{-t} \frac{1}{1-u_{\kv,\omega} t},
\label{eq:fullGreen}
\end{align}
where $u_{\kv,\omega} = \bbt (U/2)^2 G(\kv,\omega) G(\kv+\Qv,\omega)$.
 Solving this equation, we obtain the spectral function
\be
A(\kv,\omega)=\frac{\abs{\bw+\emi^{-}_\kv}}{\mathbb{t} (U/2)^2}\exp\left[-\frac{\bw^2-(\emi^{-}_\kv)^2}{\mathbb{t}
(U/2)^2}\right] \Theta(\bw^2-(\emi^{-}_\kv)^2),
\label{ll}
\ee
where, we remind, $\bar{\omega}=\omega+\mu-\varepsilon^{+}_\kv$. This spectral function has two unknowns:
the chemical potential $\mu(x, T)$ and the spin correlation length $\xi(x, T)$, which appears in $\bbt \sim \ln \xi$.
One
  condition on $\mu$ and $\xi$  is
  the constraint on the fermion density
 \begin{widetext}
\begin{align}
\frac{1-x}{2}=\int \diff \omega \int \frac{\diff \kv}{(2\pi)^2} A(\kv,\omega) n_F(\omega)
=\int \diff \omega \int \frac{\diff \kv}{(2\pi)^2} \frac{\abs{\bw+\emi^{-}_\kv}}{\mathbb{t}
(U/2)^2}\exp\left[-\frac{\bw^2-(\emi^{-}_\kv)^2}{\mathbb{t} (U/2)^2}\right] \Theta(\bw^2-(\emi^{-}_\kv)^2)n_F(\omega).
\label{eq:selfconsistentDis1}
\end{align}
\end{widetext}
To get the other condition,
we relate $\xi$ to the particle-hole polarization bubble in the same way as in
  the one-loop formula,
$\xi = (A/(1-U \Pi^{(s)} ({\bf Q},0)))^{1/2}$ (see Eq. (\ref{mm}) in Sec. \ref{sec:oneloopdisorder}), but including
the series of thermal corrections into  $\Pi^{(s)} ({\bf Q},0)$.

\begin{figure}[h]
\includegraphics[width=1\columnwidth]{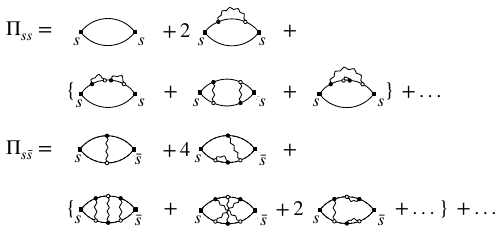}
\caption{Diagrammatic series for the spin polarization bubble.
The quantity we need is  $\Pi^{(s)}(\Qv,0)=\Pi_{ss}(\Qv,0)-\Pi_{s \bar{s}}(\Qv,0)$,
 where  $s = \uparrow$ and ${\bar s}  = \downarrow$ are spin components at the two side vertices. \label{fig:bubbleinfty}}
\end{figure}

To evaluate
 $\Pi^{(s)}(\Qv, 0)$, we note that both vertex corrections and corrections to the fermion Green's function should be
 included on an equal footing.
   The spin structure of the electron-magnon coupling implies that  $\Pi^{(s)}(\Qv, 0)=\Pi_{zz} (\Qv,0)=\Pi_{s s}
   (\Qv,0)-\Pi_{s \bar{s}}(\Qv,0)$, where $\Pi_{ss'}$ denotes the bubble diagram with spin indices $s$ and $s'$ at
   the two side vertices, and $s = \uparrow$, ${\bar s}  = \downarrow$.
We show the corresponding diagrams in Fig.~\ref{fig:bubbleinfty}.
 We find (see~\cite{SM}
 for details)
\begin{align}
\Pi^{(s)}
(\Qv,0) = &
\frac{-1}{\bbt (U/2)^2} \int \diff \omega \int \frac{\diff \kv}{(2\pi)^2} \Bigg( n_F(\omega) \sgn(\bar{\omega})\times
\non\\
&\Theta(\bar{\omega}-(\emi^{-}_\kv)^2) \exp\left(-\frac{\bar{\omega}^2 - (\emi^{-}_\kv)^2 }{\bbt
(U/2)^2}\right)\Bigg).
\label{eq:Pizz}
\end{align}

Note that
  because we included the same number $n=2$ of fluctuating spin modes in the SDW and the paramagnetic state,
   the condition $\xi^{-1}=0$, i.e., $\Pi^{(s)} (\Qv,0) =1/U$, is equivalent to the condition on $\Delta =0^+$ in Eq.
   ~\eqref{eq:selfconsistentSDW2}.

We present the results in the next section. Before that, two
    comments are in order.
   First,
    the eikonal
   approach,
    which we  employ here,  is valid when the momentum integration in each diagram can be fully confined to the bosonic propagator. This holds
  when the pseudogap energy $\Dpg^{(e)}$ (the half-distance between the humps in the full spectral function) found from the eikonal approach exceeds
    $v_F \xi^{-1}$.  Once this condition breaks down,  one can no longer pull out the Green's functions from the
    momentum integrals.  In this situation,  the eikonal approach becomes  uncontrollable.
    We will discuss this in more detail in the next section.
    Second, we re-iterate that for the fully self-consistent analysis,
      one should include longitudinal spin fluctuations across $T_N$. To account for  these fluctuations in the SDW
      state, one would need to
      introduce another tunable parameter $\log {\delta_\text{amp}}$ for the amplitude mode.

We also note that Eqs.~\eqref{kk},~\eqref{eq:selfconsistentSDW1} and~\eqref{eq:selfconsistentSDW2} below $T_N$ and Eqs.~\eqref{ll},~\eqref{eq:selfconsistentDis1} and~\eqref{eq:Pizz} above $T_N$ can be reproduced in the path-integral analysis, which maps the eikonal approximation onto the annealed disorder problem, similar to the discussions in Refs.~\cite{Schmalian1999,Berg2007,Sadovskii_review}.  We discuss path-integral approach in the Supplementary Material~\cite{SM}.

\section{Results and Discussions}
\label{sec:EikonalResults}

In this section, we present the solutions of  Eqs.~\eqref{kk},~\eqref{eq:selfconsistentSDW1} and~\eqref{eq:selfconsistentSDW2} below $T_N$ and Eqs.~\eqref{ll},~\eqref{eq:selfconsistentDis1} and~\eqref{eq:Pizz} above $T_N$ at the hole doping $x=0.05$, for different values of the Hubbard $U$.

We present the results for (i) the spectral function $A(\omega)$, (ii) the spectral    intensity    ${\bar A} (\omega) = A(\omega) n_F (\omega)$, proportional to photoemission intensity, (iii) the SDW order parameter $\Delta$, (iv) the pseudogap energy $\Dpg^{(e)}$, which is the half-distance between the two peaks in the spectral function, (v) the energy scale $v_F \xi^{-1}$ associated with magnetic fluctuations, (vi) the temperature dependence of the  coupling constant $\lambda (T) \propto T/(v_F \xi^{-1})^2$, and (vii) the change of chemical potential, $\delta \mu = \mu - \mu_0$, we determine self-consistently from the condition that the total fermionic density equals $1-x$.

\begin{figure}[h]
\subfigure[]{\includegraphics[width=0.48\columnwidth]{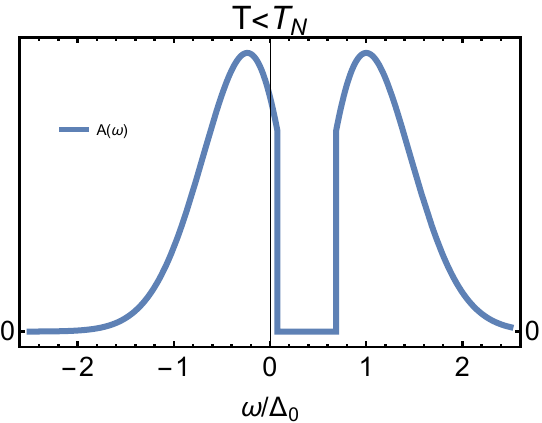}}
\subfigure[]{\includegraphics[width=0.48\columnwidth]{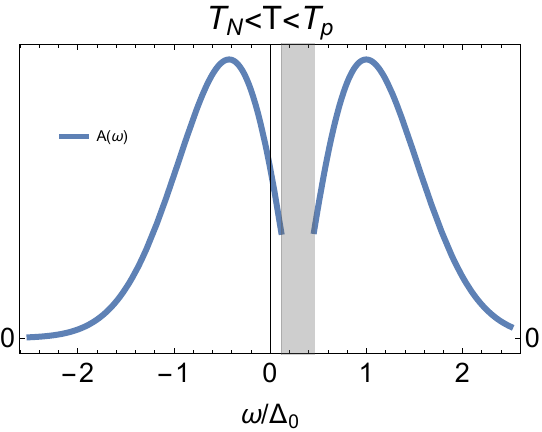}}
\subfigure[]{\includegraphics[width=0.48\columnwidth]{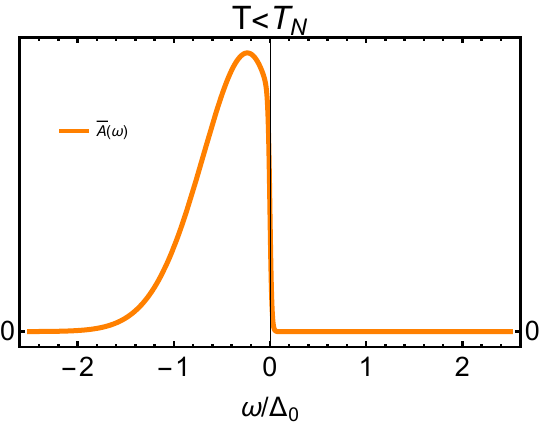}}
\subfigure[]{\includegraphics[width=0.48\columnwidth]{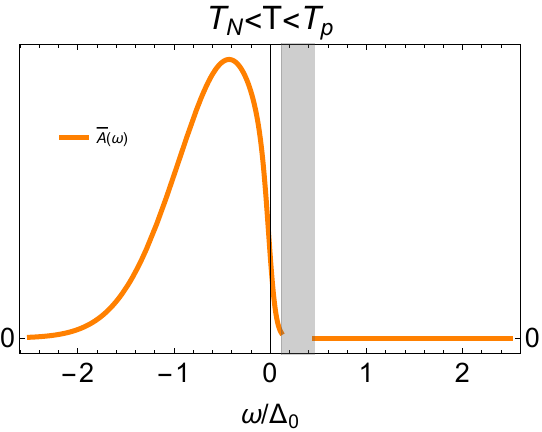}}
\caption{Spectral function $A(\omega)$ (blue line in the upper pannel) and spectral intensity $\bar{A}(\omega)$ (orange line in the lower pannel)  in (a,c) the SDW state for $T=0.01\eV$,
 from Eq.~\eqref{kk},
and (b,d) the pseudogap regime for $T=0.04\eV$, from Eq.~\eqref{ll}. We set $x=0.05, U=2\eV$, $\epsilon=0.01$ and $T_N=0.013\eV$.
   The gray shaded area in (b) is the region $\abs{\omega+\delta \mu}<v_F \xi^{-1}$, where the eikonal  approach
 breaks down.
\label{fig:spectralnumerical}}
\end{figure}

In Fig. \ref{fig:spectralnumerical} we show the spectral function $A(\omega)$ (upper pannel) and the spectral intensity $ {\bar A} (\omega)$ (lower pannel) for  $U=2\eV$. The left and right panels show the results in the SDW state and the paramagnetic state, respectively. The spectral function in the SDW state has a true gap $\Delta$ and two maxima separated by $2\Dpg^{(e)} \sim U$. The spectral intensity in both cases shows only a maximum (a pseudogap) at the energy $\omega = -(\Dpg^{(e)} +\delta \mu)$.  At large $U$, $\Dpg^{(e)} \approx U/2$ and $\delta \mu \approx -U/2$, the sum is of order $J$. At intermediate $U=2eV$, the frequency where $ {\bar A} (\omega)$ has a maximum is comparable with $\Delta_0$. The shaded region in the paramagnetic state marks the condition $|\omega+\delta \mu| < v_F \xi^{-1}$, for which the eikonal approach is not applicable.

\begin{figure*}[ht]
\subfigure[]{\includegraphics[width=0.6\columnwidth]{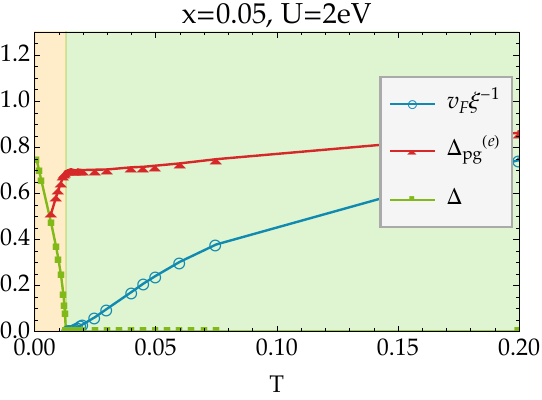}}\quad
\subfigure[]{\includegraphics[width=0.6\columnwidth]{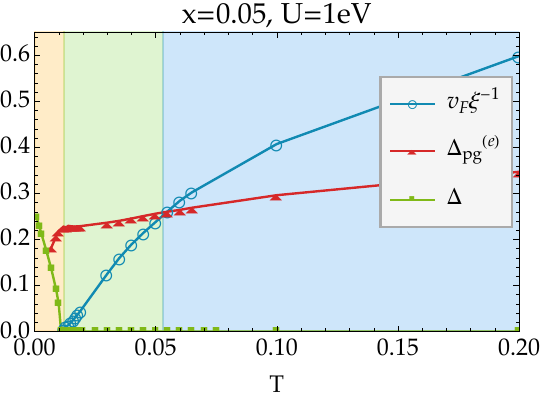}}\quad
\subfigure[]{\includegraphics[width=0.58\columnwidth]{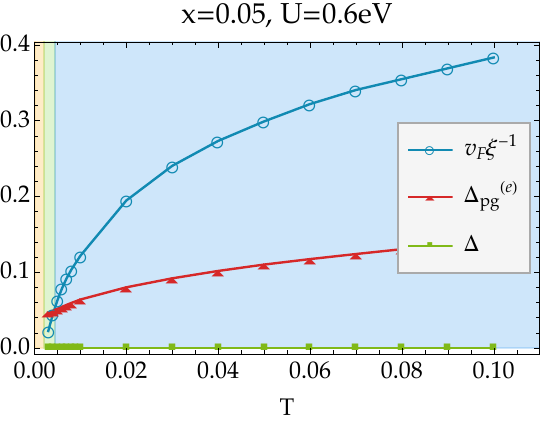}}
\subfigure[]{\includegraphics[width=0.6\columnwidth]{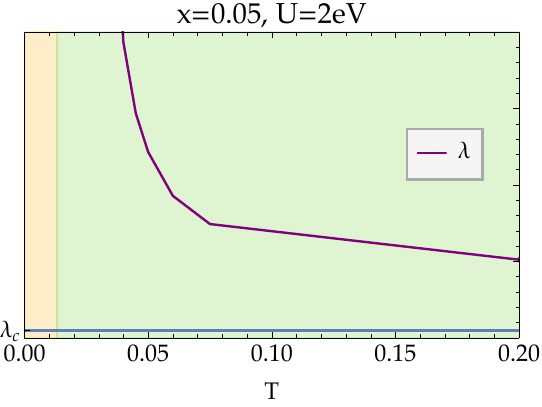}}\quad
\subfigure[]{\includegraphics[width=0.6\columnwidth]{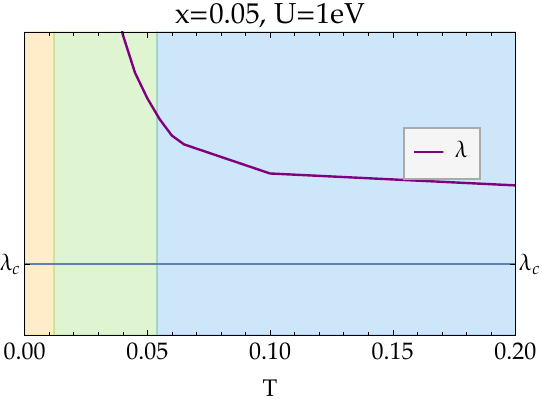}}\quad
\subfigure[]{\includegraphics[width=0.58\columnwidth]{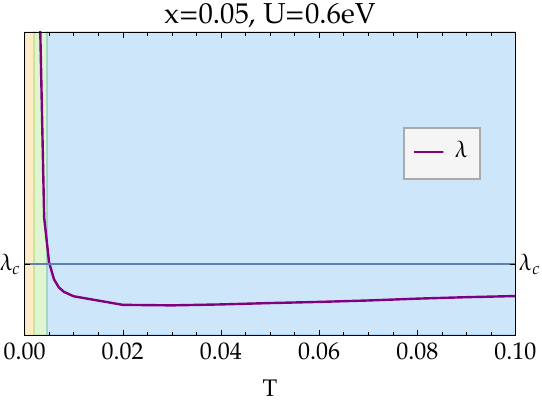}}
\subfigure[]{\includegraphics[width=0.6\columnwidth]{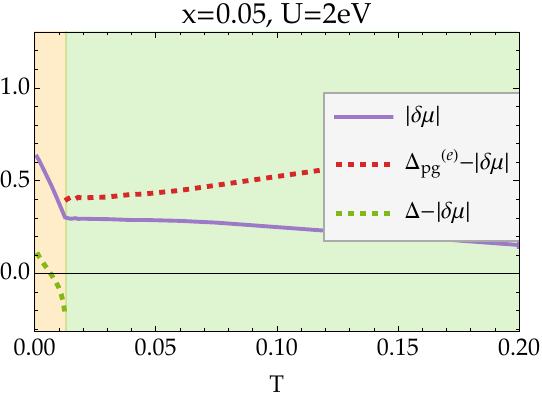}}\quad
\subfigure[]{\includegraphics[width=0.6\columnwidth]{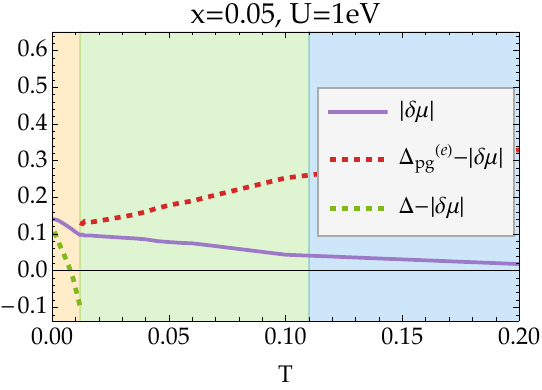}}\quad
\subfigure[]{\includegraphics[width=0.58\columnwidth]{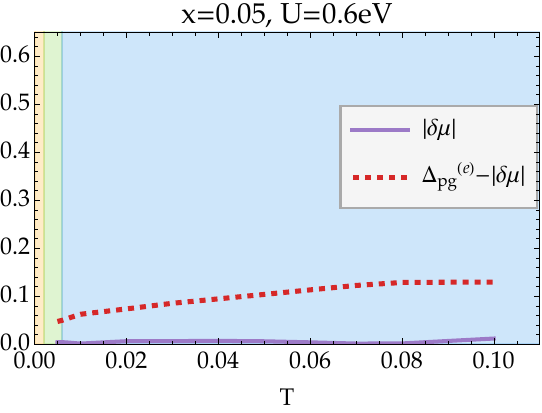}}
\caption{Results for $U=2.0\eV$ (left column), $U=1.0\eV$ (middle column) and $U=0.6\eV$ (right column). We used $t=0.3\eV, t'=-0.06,  x=0.05, \epsilon=0.01$. (a)-(c): Temperature evolution of the energy scales (in $\eV$) $\Delta$, $\Dpg^{(e)}$ and $v_F \xi^{-1}$.  (d)-(f): Temperature evolution of $\lambda\propto T \xi^{2}$, which determines the behavior of the one-loop pseudogap $\tDpg$. Note that the numerical prefactors for $\lambda$ are not taken into account in the plots.
  (g)-(i): Temperature evolution of the shifted chemical potential due to interaction, $\delta \mu=\mu-\mu_0$. The
 distances of the pseudogap and SDW gap from the chemical potential  are plotted as $\Dpg^{(e)}-|\delta \mu|$ and
 $\Delta-|\delta \mu|$. Yellow, green, and blue areas denote, respectively, the SDW state, paramagnetic state with
 strong pseudogap behavior, and paramagnetic state with weak pseudogap behavior.  The results in the blue shaded area are approximate
  as the eikonal approach breaks down in this region (see the text).
\label{fig:Escales}}
\end{figure*}

In Fig.~\ref{fig:Escales} we show how the different characteristic energies vary with temperature.
 The results  for $U=2\eV$ and $U=1\eV$, for which the ground state is SDW ordered over a wide range of $T$,
  are shown in the left and middle columns.  The bandwidth is set to be
 $W=8 t = 2.4 \eV$.
 In panels (a-b) we show the
temperature evolution of
 $\Delta$, $\Dpg^{(e)}$ and $v_F \xi^{-1}$
  and in panels (d-e) we show the temperature dependence of the one-loop coupling constant $\lambda \propto T \xi^2$ (purple line) with $\xi (T)$, obtained by summing up the eikonal series.

  We see that the pseudogap does develop in the SDW state, and the pseudogap energy
  $\Dpg^{(e)}$ increases with $T$, roughly as $\sqrt{T}$, while the SDW order parameter $\Delta$ decreases with $T$ and
  vanishes at $T_N$.
   Within our numerical accuracy the pseudogap
  becomes visible above a small, but finite $T$. Analytically, we found  that it  develops already at infinitesimally $T$.

\begin{figure}[h]
\includegraphics[width=0.8\columnwidth]{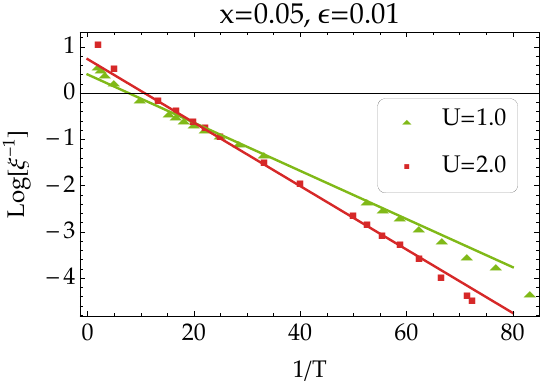}
\caption{Spin correlation length $\xi$ as a function of temperature $T$. Over a temperature range (above $T_N$ and
when $\xi(T)\lesssim1$), we find the fitting functions $\xi^{-1}(T)=1.86 \eu^{-\frac{0.06}{T-0.0015}}$ for $U=2\eV$
and $\xi^{-1}(T)=1.26 \eu^{-\frac{0.04}{T-0.0027}}$ for $U=1\eV$. The deviation from the 2D $n=3$ non-linear sigma
model (NLSM) behavior
 $\xi^{-1} \sim \eu^{-\rho_s/T}$ in the denominator of the exponential factor is due to the cutoff $\epsilon$
 introduced in the computation, which is not relevant when $\epsilon<\xi^{-1}$ at $T>T_N$. At high temperature when
 $\xi\sim \mathcal{O}(1)$, the numerical results also deviate from the 2D
NLSM behavior.
\label{fig:logxi}}
\end{figure}

  In the paramagnetic phase above $T_N$, we find that the pseudogap energy is quite flat (more so for larger $U =2 \text{eV}$).
  We call this a strong pseudogap behavior.
   Because the pseudogap energy $\Dpg^{(e)} \sim \sqrt{T \log{\xi}}$, the near temperature independence of $\Dpg^{(e)}$ implies that
   the fully renormalized $\xi$ decreases rapidly, almost exponentially with $1/T$.  In Fig.~\ref{fig:logxi} we plot
   $\log{\xi}$ as a function of the inverse temperature. We see that the temperature evolution of $\log{\xi}$ is
   indeed nearly linear in $1/T$, i.e.,
    to a reasonable accuracy, $\xi \sim \eu^{T_0/T}$.  We emphasize that this result is obtained by
    computing $\xi$  self-consistently from the four-point correlation function in the metallic state,
     which we compute by including self-energy and vertex corrections due to  thermal
      magnetic
    fluctuations to all orders in perturbation theory.
        The exponential behavior mimics
    the one in the non-linear sigma model of localized
    spins~\cite{Polyakov1975,Nelson1989,Hasenfratz1991,SachdevBook}, but we emphasize that we found this behavior in a metal.

   For $U=2\text{eV}$ the pseudogap energy $\Dpg^{(e)}$ remains larger than $v_F \xi^{-1}$ over the whole $T$ range covered in Fig. \ref{fig:Escales},
    and in most of this range the system displays  a strong pseudogap behavior.

     For smaller $U=1\eV$, the condition $\Dpg^{(e)}$ remains larger than $v_F \xi^{-1}$
      up to a certain $T_{\rm{cross}} > T_N$ (Fig.~\ref{fig:Escales}~b).
       At larger $T$, the eikonal approach
     breaks down. To understand the behavior at these $T$, we note that
     the actual pseudogap energy, $\Dpg$, obtained from the full Green's function, is comparable
      to one-loop pseudogap energy $\tDpg$.  The ratio of the latter and $v_F \xi^{-1}$ is controlled by the parameter $\lambda$, defined in (\ref{zz_3}) and is large when $\lambda$ is large.
     Accordingly, the  eikonal approach is valid when $\lambda \gg 1$.
       We plot $\lambda (T) \propto T \xi^2 (T)$ in Fig.~\ref{fig:Escales} (d, e), using $\xi (T)$ extracted from the fully dressed polarization bubble.  We see that $\lambda$ is indeed large when  $\Dpg > v_F \xi^{-1}$.
       It diverges at the boundary of the SDW order where $\xi (T)$ diverges.

\begin{figure}[h]
\subfigure[]{\includegraphics[width=0.48\columnwidth]{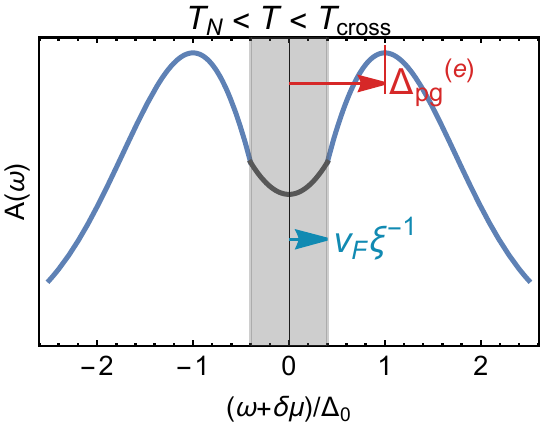}}
\subfigure[]{\includegraphics[width=0.48\columnwidth]{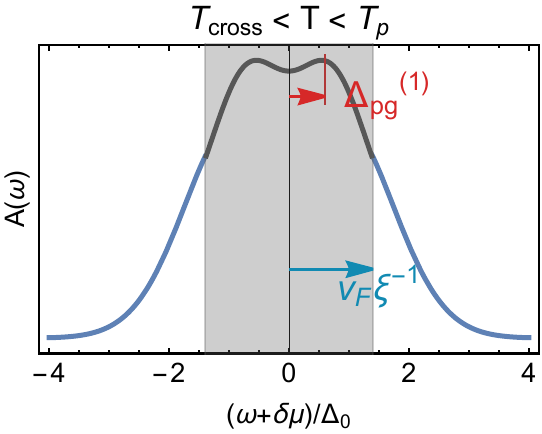}}
\caption{
The spectral function
  in the (a) strong pseudogap regime ($\Dpg^{(e)} > v_F \xi^{-1}$) and (b) weak pseudogap regime ($\Dpg^{(e)} < v_F \xi^{-1}$),
    where $\Dpg^{(e)}$ is the pseudogap obtained within the eikonal approach by summing up an infinite series of thermal
    contributions to the self-energy (blue solid line).
   The gray shaded area is the frequency range where the eikonal approach
   breaks down, and one
   should instead use the full one-loop result for the self-energy (gray solid line).
    $\tDpg$ is the pseudogap energy from such a one-loop calculation.
\label{fig:spectralill}}
\end{figure}

          At $\lambda = O(1)$, we expect the one-loop expression for the self-energy to be sufficient, at least for qualitative reasoning. The one-loop pseudogap still exists at $\lambda \geq 1$, but the pseudogap energy $\tDpg$ decreases with decreasing $\lambda$  and vanishes
    at $\lambda = \lambda_c =0.47$ (see Fig.~\ref{fig:schematicE}). We  follow~\cite{Schmalian1998} and call this
    a weak pseudogap regime.
    In Fig.~\ref{fig:spectralill}, we show the spectral function in both regimes. How far in $T$ a weak pseudogap behavior extends depends on temperature
     variation of $\lambda$ in the blue region
      in Fig.~\ref{fig:Escales}~(e),  where the eikonal approach is no longer controllable, and the temperature variation of $\xi$ cannot be obtained rigorously.
     Yet, we see from  Fig.~\ref{fig:Escales}~(e) that $\lambda$ still strongly decreases with $T$
      when $\Dpg^{(e)}$ and $v_F \xi^{-1}$ become comparable.  It is then natural to assume that it continues decreasing at higher $T$ and reaches
       critical $\lambda_c$ at some $T_p$.  We expect a weak pseudogap behavior  to exist also for $U=2 \eV$, but at temperatures higher than the ones that we
           probe numerically.

Overall, for the values of $U$, for which the ground state is ordered, the pseudogap develops inside the SDW state, remains finite at $T_N$, persists into the paramagnetic phase and remains weakly $T$ dependent up to $T_{\rm{cross}}> T_N$. It then decreases with increasing $T$ and eventually vanishes at $T = T_p$ (see Fig.~\ref{fig:schematicE}). This behavior is quite consistent with the results of several numerical studies~~\cite{Schafer2021}.

In  panels (g-h) we show for these $U$
the temperature evolution of the  shift of the chemical potential $\delta \mu = \mu -
\mu_0$, where, we remind, $\mu$ is the actual chemical potential and  $\mu_0$ is the chemical potential for free fermions. We emphasize that $\delta \mu$ is  negative in the
whole temperature range,
    where our approach is valid.
    In the same panels we plot
    $\Dpg^{(e)}-|\delta \mu|$ and $\Delta-|\delta \mu|$.  When the difference is positive (which is the case for
    $\Dpg^{(e)}-|\delta \mu|$ for all $T$ and for  $\Delta-|\delta \mu|$ at low $T$),  the spectral intensity ${\bar A}
    (\omega)$ at a hot spot
     has peaks  at a negative $\omega$, where $|\omega| = \Dpg^{(e)} -|\delta \mu|$ and $|\omega| =  \Delta-|\delta \mu|$.

\begin{figure}[t]
\includegraphics[width=0.9\columnwidth]{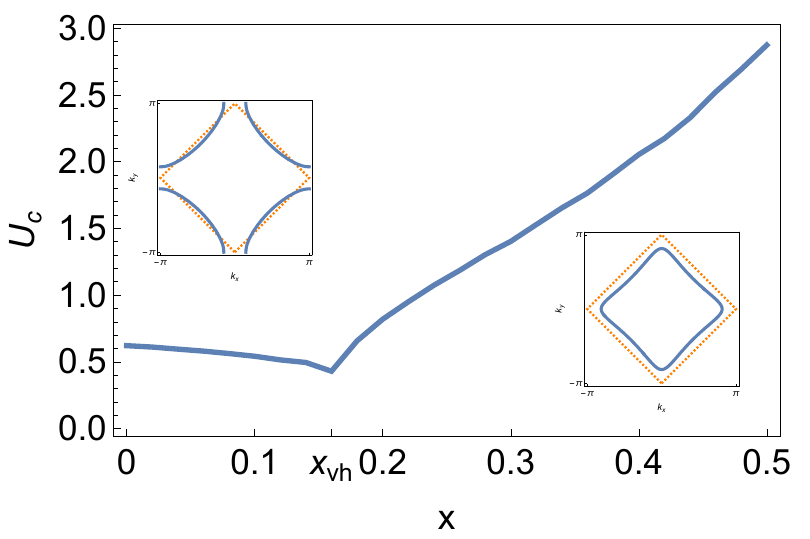}
\caption{Threshold $U_c$ above which the $(\pi, \pi)-$SDW is a mean field solution at different doping $x$. $U_c$ is the smallest at $x_{vh}$ at which the Fermi surface changes topology. See figure insets for the non-interacting Fermi surface (dark blue line) below $x_{vh}$ (left inset) and above $x_{vh}$ (right inset).
\label{fig:Ucx}}
\end{figure}
We next discuss what happens when the ground state is not ordered.  We plot critical $U_c (x)$, at which the SDW order disappears at $T=0$ at a given $x$,  in Fig.~\ref{fig:Ucx}.  Note that this is the mean-field value of $U_c$ as we consider only thermal fluctuations.
 For $x=0.05$, $U_c =0.58$.  In panels (c,g,i) in Fig.~\ref{fig:Escales} we present the results for
$x=0.05$ and  $U=0.6\text{eV}$, which is close to $U_c$.
   In numerical calculations we find no SDW order above $T\approx 0.002\eV$ for this $U$,
 and very tiny SDW order $\Delta \approx 0.01\eV$ at $T\leq 0.002\eV$.
 Because both $\Dpg^{(e)}$, generated by thermal fluctuations, and $\xi^{-1}$ necessary vanish at the SDW QCP,  whether
   a strong pseudogap behavior
 exists at a finite $T$ right above the QCP depends on the interplay between temperature variations of $\Dpg$ and $v_F \xi^{-1}$, and whether a weak pseudogap behavior exists depend on  whether the temperature dependent coupling $\lambda$ is larger than the critical $\lambda_c$.  We see from Fig.~\ref{fig:Escales} (c) that $v_F \xi^{-1} > \Dpg$ at all $T$, except the very lowest.  We decreased $U$ to $U_c$ and to $0.57\eV < U_c$ and verified that within our numerical accuracy,  $\Dpg < v_F \xi^{-1}$ for all $T$.   This implies that strong pseudogap behavior does not develop if there is no SDW order at $T=0$.  Furthermore, we see from Fig.~\ref{fig:Escales} (f) that $\lambda$ remains smaller than $\lambda_c$ for all  $T > 0.002\eV$.
  We verified that at
  $U=U_c$, $\lambda<\lambda_c$ at all $T$ within our numerical reach (see Fig.~\ref{fig:U0d58}~b and discussions in Sec.~\ref{sec:phase_diagram}). Hence, a weak pseudogap behavior also does not develop if the ground state is not SDW-ordered.
 In other words,
    the pseudogap behavior  holds only above the SDW ordering temperature $T_N (x)$, but does not extend to dopings, for which $T_N=0$.

 It is instructive to compare our results with the ones by
 Schmalian, Pines, and Stojkovi\'c (SPS) (Ref.~\cite{Schmalian1999}), who also studied the evolution of the spectral function in the paramagnetic state in a non-perturbative fashion
 (see also Ref. \cite{Sadovskii_review, Sadovskii1999}).
        SPS assumed that the static magnetic susceptibility can be factorized as
        \be
        \chi(\tilde{\qv}+\Qv, 0)\sim \frac{1}{\xi^{-2}+\tilde{q}_{\parallel}^2 + \tilde{q}_{\perp}^2}\Rightarrow
        \frac{\xi^{-1}}{\xi^{-2}+\tilde{q}_{\parallel}^2 } \frac{\xi^{-1}}{{\xi^{-2}+ \tilde{q}_{\perp}^2}}.
          \label{pp}
          \ee
          This allowed them to obtain the
       iterative equation for the fermion Green's function between $j$th and $(j+1)$th loop orders
         and sum up the contributions from all loop orders.
        In the limit $v_F \xi^{-1}\ll \Delta_{pg}$, their and our approaches yield the same diagrammatic series for
        $G$, whereas
         for $v_F \xi^{-1} \gg \Delta_{pg}$,  the two results agree up to a numerical factor.
         The advantage of the SPS approach, based on (\ref{pp}), is in that it allows one to analyze analytically the
         crossover between the strong and weak pseudogap regimes. The disadvantage is that it does not allow one to connect to  pseudogap
         behavior in the SDW phase, because when $\xi \to \infty$ it yields $\int d^2 \qv \chi(\tilde{\qv}+\Qv, 0) =
         O(1)$ instead of divergent $\int d^2 \qv \chi(\tilde{\qv}+\Qv, 0) \sim \log {\xi}$, which we obtained
         without factorization.
          We also note that SPS took $\xi$ as an input parameter, while we
        compute it self-consistently, in the same eikonal-type approach. This
        is essential for the
        understanding of the temperature evolution of $\Delta_{\rm pg}$ in the paramagnetic phase.
         In particular, we argue in the next section that temperature dependence of the fully dressed $\xi$ is such that
          pseudogap does not develop if the ground state is \emph{not} magnetically ordered.

\subsection{Phase diagram}
\label{sec:phase_diagram}
To convert our results
into the phase diagram in the $(T,x)$ plane,
 we need to locate the parameter range where the thermal contribution
to the self-energy is larger than the combined contribution from non-zero bosonic Matsubara frequencies.
  For systems with localized spins, there is no such regime as $\xi^{-1}$
 is linear in $T$, and for typical momenta $q \sim \xi^{-1}$, the static part of the inverse bosonic propagator $\xi^{-2} +q^2$  has the same $T^2$ temperature dependence as the dynamical $\omega^2_m \sim T^2$ term.  Then
 thermal and quantum  fluctuations are comparable in strength in the whole low-energy range above a QCP.
 The  phase diagram contains an ordered phase, a renormalized classical phase adjacent to it, a quantum-critical phase,
 where $\xi^{-1} \propto T$, and a quantum-disordered phase~\cite{SachdevBook}.
  For metals with dynamical exponent $z=1$ (the case when Landau damping of critical fluctuations is absent by kinematic reasons) the phase diagram is similar, with an extra region of Fermi-liquid phase on the paramagnetic side of the QCP.

For metals with $z >1$,  the static part of the inverse bosonic propagator scales as $\xi^{-2}$, for typical momenta $q \sim \xi^{-1}$, while the dynamical part scales as $T^{2/z}$. The two $T$ dependencies are generally different, even if
$\xi^{-1} \sim T$. At small $T$, it is natural to expect that  $\xi^{-2}$ is smaller than properly normalized  $T^{2/z}$.
 Then thermal fluctuations give the largest contribution to the self-energy.
  As $T$ increases, this condition $\xi^{-2} < T^{2/z}$  may or may not
hold, depending on the thermal evolution of $\xi$. If it holds for all
$T$,
 where the low-energy description is applicable, thermal fluctuations completely determine system behavior
 above a QCP.
 If it breaks down at some $T= T_q$ within the low-energy regime,
 then at this temperature
the system crosses over
from
  thermal fluctuations dominated non-Fermi liquid behavior at $T < T_q$ to
   still non-Fermi liquid behavior, but with the largest contribution to the self-energy coming from the terms with a non-zero bosonic
    Matsubara frequency.

 This reasoning holds when in the  thermal regime the system displays a pseudogap behavior (strong or weak).
   In our notations, this implies that $T_q$ must be smaller, or, at most, compatible to $T_p$. If $T_q > T_p$,
  a separate consideration is required for the region $T_p < T < T_q$. 

 To compare the two temperatures, we note that $T_p$ corresponds to $\lambda = \lambda_c$. Using the definition of $\lambda = \lambda (T)$, Eq. (\ref{zz_3}), we find that $T_p$ is the solution of  $T_p \xi^2 (T_p)\sim v^2_F /{\bar g}$.
 The temperature $T_q$ is determined by comparing the Landau damping term
 at $\omega \sim T$ and $q \sim \xi^{-1}$ to $\xi^{-2}$.  The Landau damping of spin excitations comes from scattering
  into low-energy fermions, and the effective coupling for this process is the same ${\bar g}$ as in Eq. (\ref{ii}) for the self-energy.  Evaluating the Landau damping term, we find that the equation on $T_q$ is, up to a numerical factor, the same as for $T_p$: $T_q \xi^2 (T_q)\sim v^2_F /{\bar g}$. Then $T_q$ and $T_p$ are comparable, i.e.,
   the thermal region is also the pseudogap region.

\begin{figure}
\subfigure[]{\includegraphics[width=0.48\columnwidth]{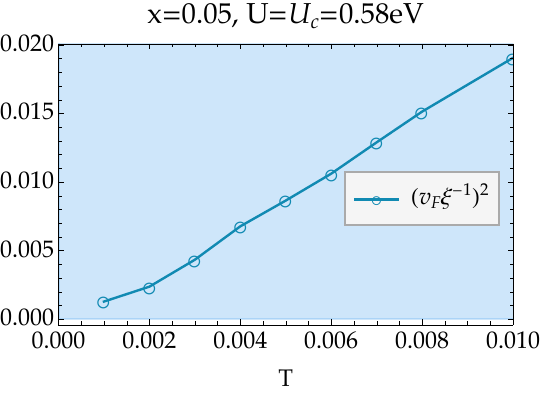}}
\subfigure[]{\includegraphics[width=0.48\columnwidth]{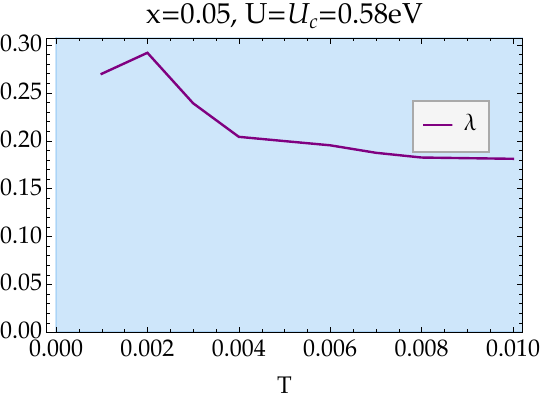}}
\caption{(a) $(v_F \xi)^{-2}$ (b) $\lambda \propto T \xi^{2}$ as
 functions
 of temperature at the critical $U_c=0.58\eV$ for doping $x=0.05$.
\label{fig:U0d58}}
\end{figure}

 Whether $T_p$ is finite right above the SDW QCP depends on the temperature variation of  $\xi^{-2}(T)$.
 If this variation was analytic  $\xi^{-2}(T) \sim T^2$,  the coupling $\lambda \propto T \xi^2 (T)$ would necessarily be large at small $T$, and hence $T_p$ would be finite. In this situation the pseudogap region would extend into the doping range where the ground state is not magnetically ordered.  We find however, that above the SDW QCP, $\xi^{-2} \propto T$ (see Fig.~\ref{fig:U0d58}~a)  In this case, $\lambda$ becomes $T$ independent, and  pseudogap develops if this constant $\lambda$ is larger than $\lambda_c$ and does not develop if it is smaller.
  As we already said,
  our results show that $\lambda < \lambda_c$ (see Fig.~\ref{fig:U0d58}~b), hence pseudogap does not
  develop right above the QCP.
     By continuity, it also
     does not exists  in the range where the ground state is not magnetically ordered.
     We caution, however, that this result is likely model-dependent, and
     in a more generic model with a non-local interaction
   the magnitude of $\lambda$ above a QCP may exceed $\lambda_c$.
      In such a case the pseudogap extends into the range where the ground state is not ordered.

 The linear in $T$ behavior of $\xi^{-2}$  above a QCP (modulo $\log T$) holds in  the Hertz-Millis theory of the SDW quantum criticality~\cite{Millis1993},  where it appears due to theremal bosonic self-energy from the phenomenologically introduced mode-mode coupling. In a microscopic theory,  mode-mode coupling appears  as an effective 4-boson interaction -- a square made out of four fermionic propagators.  The contribution from this effective interaction
     to bosonic self-energy is the same one as from inserting self-energy and vertex corrections into the polarization bubble~\cite{Abanov2003}, which are elements of our diagrammatic series for the fully dressed $\xi$. Not surprisingly then, we obtain the same linear in $T$ dependence of $\xi^{-2}$ as in Hertz-Millis theory.
     We note, however, that we collected infinite number of graphs for the bosonic propagator, not only the lowest order vertex and self-energy corrections.

We are now in position to obtain the phase diagram  in variables $x$ and $T$. For this, we combine our results (i) that there is no pseudogap at $T=0$ from dynamical fluctuations, (ii) that the pseudogap behavior at a finite $T$ exists only for $x < x_c$, when there is a SDW order in the ground state, (iii) that there is a quantum-critical behavior at $T > T_p$, and (iv) that at $x > x_c$ there is a Fermi liquid behavior at the smallest $T$. We show the phase diagram in Fig. \ref{fig:schematicE}.

In the two right panels in Fig.~\ref{fig:PhaseDiagramQT} we present the phase diagrams for the cuprates, suggested by our study.
In both figures pseudogap behavior due to thermal magnetic fluctuations exists only when the system has a SDW $(\pi,\pi)$ order at $T=0$.  In Fig.~\ref{fig:PhaseDiagramQT} (b) this range is narrow and ends well before optimal doping.
This phase diagram is likely applicable to  hole-doped cuprates, where SDW region is quite narrow. We argue therefore  that the pseudogap behavior, observed in these systems, is not caused by thermal spin fluctuations and is instead either a precursor to superconductivity~\cite{Randeria1998,Millis1998,Berg2007,YMWu2021,Qi2022},
 or a novel state of matter with  current-like or topological order~\cite{Varma1997,Varma1999, Sachdev2018,YHZhang2020a,YHZhang2020b,Sachdev2022pga,Sachdev2022pgb}.  In the phase diagram in Fig. \ref{fig:PhaseDiagramQT}~c, the
 range of SDW order is wider and extends to near-optimal doping. In this case pseudogap behavior due to thermal magnetic fluctuations exists in a wider parameter range,
 and $T_p$, up to which it holds, may be close to the experimental boundary of the pseudogap phase.
 This last behavior holds in electron-doped cuprates~\cite{ArmitageRMP}, and we believe that pseudogap behavior, observed in these materials, may actually be due to thermal spin fluctuations.

  A comment is in order here.  In this paper we restricted our analysis to $(\pi,\pi)$ SDW order.
   The recent numerical study~\cite{Simkovic2022} suggested that
    the pseudogap may exist as long as the ground state has a stripe magnetic order, which can be viewed as partly melted
  incommensurate $(\pi, Q)/(Q,\pi)$ SDW order~\cite{Shraiman_1989,Dombre1990,ChubukovMusaelian1994,Zhou1995,White2021,Arovas2022,Simkovic2022}.
  Such an order has been extensively studied, chiefly in La-based cuprates
   (see e.g., ~\cite{Haug_2010,Tranquada2013} and references therein).  This may potentially widen the range of magnetically induced pseudogap behavior even in hole-doped cuprates.

 \section{Summary}
\label{sec:summary}

To summarize, in this paper, we analyzed the precursor scenario for pseudogap behavior of
 interacting fermions near a magnetic instability.  We considered the Hubbard model on the square lattice and
 analyzed the thermal evolution of the spectral function.
  We adopted the eikonal-type approach, and summed  up thermal contributions to the fermion two-point and four-point
  correlation functions to infinite order both in the SDW-ordered state and in the paramagnetic state. In the latter,
  the eikonal-type computational procedure is valid at large enough magnetic correlation lengths, which we compute
  self-consistently.
  For Hubbard $U$ values comparable to the fermionic bandwidth, we found pseudogap behavior due to magnetic
  fluctuations and identified two different regimes: strong pseudogap behavior, which emerges after the summation of
  an infinite series of thermal contributions to two-point and four-point correlation functions, and weak pseudogap
  behavior, which emerges in the parameter range where the one-loop approximation is adequate.
  In the strong poseudogap regime we found that the magnetic correlation length decreases with $T$ nearly
  exponentially, as $\eu^{T_0/T}$, such that the pseudogap energy scale $\Dpg
   \propto T \log \xi$  is almost independent on $T$, despite
  that it originates from thermal fluctuations.  The near-exponential decrease of $\xi$ mimics the
    behavior in the non-linear sigma model of localized spins, but we emphasize that we obtained this behavior in a
  metal with strong correlations.
    The near-independence of the pseudogap energy on $T$ is consistent with numerical studies of the Hubbard model in the regime where the ground state is SDW-ordered (Ref.~\cite{Schafer2021}).
   At higher $T$
     and higher dopings $x$,
    we found the crossover to the weak pseudogap regime, in which the pseudogap energy gradually decreases with increasing $T$ or $x$ and eventually
     vanishes at $T_p$.

Our calculations showed that the pseudogap behavior exists only  above SDW ordering temperature  and does not extend to dopings,  where the ground state is disordered.  This in turn is consistent with quantum Monte Carlo studies of the effective  models of fermions interacting with magnetic fluctuations, as these studies didn't detect
  pseudogap behavior in the parameter range where the ground state is not magnetically ordered~\cite{BergMC1,BergMC2,BergMC3}.
 We presented in Fig.~\ref{fig:schematicE} the  phase diagram based on our model calculations  and presented
  in Fig.~\ref{fig:PhaseDiagramQT} the phase diagram for the cuprates.
  We argue that
    the magnetic pseudogap  covers the range, where the pseudogap behavior has been detected in  electron-doped cuprates, but does not cover the range of the observed pseudogap behavior in hole-doped cuprates.  The pseudogap behavior in the latter is then  either due to superconducting precursors, or is a novel ordered state.
We note, however, that we didn't analyze a potential pseudogap behavior above a stripe order~\cite{Simkovic2022}.

 The last remark.  In this paper we used the eikonal approach and set the limit of its applicability at $\Delta_{pg}^{(e)} \sim v_F \xi^{-1}$.  The eikonal approach treats vertex and self-energy corrections equally, and the applicability limit is the same for  both types of diagrams.   It is possible that there exists an intermediate regime of $\Delta_{pg}^{(e)} \sim v_F \xi^{-1}$, where self-energy corrections are numerically stronger. In this situation, one has to include
  infinite  series of self-energy corrections to the fermionic Green's function without including vertex corrections. This
   is equivalent to evaluating the Green's function in the self-consistent one-loop approximation
     (the self-energy is given by one-loop diagram, but with the full Green's function without vertex corrections of an internal fermion).
      Such an approximation has been widely used in the context of large-N limit of the SYK model and its variations~\cite{Chowdhury2022}.
    The spectral function, obtained within the self-consistent one-loop approximation gets broadened compared to the spectral function of free fermions, but  the peak of $A(k_F, \omega)$  remains at $\omega =0$, the pseudogap {\it does not} develop.
     The analysis of the interplay between the eikonal and self-consistent one-loop approximation is
     somewhat involved and will be discussed separately.

\acknowledgements
We thank Leon Balents, Erez Berg, Antoine Georges, Patrick  Lee, Izabella Lovas, Michael Sadovskii,
Subir Sachdev, J\"org Schmalian, Fedor Simkovic and particularly
 Andr\'e-Marie Tremblay for helpful discussions and suggestions. M.Y.\ was supported by the Gordon
and Betty Moore Foundation through Grant GBMF8690 to UCSB, by a grant from the Simons Foundation (216179, LB), and by
the National Science Foundation under Grant No.\ NSF PHY-1748958. ZW and RMF were supported by the Department of
Energy through the University of Minnesota Center for
Quantum Materials, under Grant No. DE-SC-0016371.  AVC  was supported by the U.S. Department of Energy, Office of
Science, Basic Energy Sciences, under Award No.\ DE-SC0014402.

\bibliography{PG_thermal}

\clearpage
\onecolumngrid
\appendix
\section{Path integral description}
\label{app:PathIntegral}
Here, we demonstrate the path integral representation for the fully renormalized fermion Green's function and spin
susceptibility with \emph{only} static magnetic fluctuations. We first review the path integral formulation in the
quasi-static limit and then obtain the expressions for the two-point and four-point correlation functions. The point
of departure is the effective action for the spin-fermion model,
\begin{align}
\mathcal{Z} &= \int \mathcal{D} \left[\bar{\psi}, \psi\right] \eu^{- \mathcal{S}_{sf}},\non\\
\mathcal{S}_{sf} &= \int_\tau \int_{\tau'} \left(\sum_{\kv, s} \bar{\psi}_{\kv, s} (\tau) \left(-
\mathcal{G}_{0,\kv}^{-1}\psi_{\kv,s} (\tau')\right) - 2g^2 \sum_\qv \chi_{\qv} (\tau - \tau')
\hat{\vec{S}}_\qv (\tau)\cdot\hat{\vec{S}}_{-\qv} (\tau')\right) ,
\end{align}
where $\int_\tau = \int_0^{\beta} \diff \tau$, $\mathcal{G}^{-1}_{0,\kv}(\tau-\tau') = -(\partial_\tau +
\epsilon_\kv-\mu) \delta(\tau-\tau')$, $\hat{\vec{S}}_\qv (\tau) = \frac{1}{2}\sum_\kv \bar{\psi}_{\kv+ \qv,s} (\tau)
\hat{\vec{\sigma}}_{s,s'} \psi_{\kv,s'}(\tau)$. Here, $\psi$ is the Grassmann fermionic field. For now, we define $g$
as a phenomenological coupling constant. In the Matsubara frequency representation, the action becomes
\begin{align}
\mathcal{S}_{sf} = \sum_{\kv, \sigma, n} \bar{\psi}_{\kv, \sigma, n} (-\iu \omega_n + \epsilon_\kv-\mu ) \psi_{\kv,
\sigma, n} - 2g^2 \beta^{-1} \sum_{\qv,m} \chi_\qv (\Omega_m) \hat{\vec{S}}_\qv (\Omega_m) \cdot \hat{\vec{S}}_{-\qv}
(-\Omega_m)
\end{align}
where $\beta=1/T$, $\psi(\tau) = \frac{1}{\sqrt{\beta}} \sum_n \eu^{-\iu \omega_n \tau} \psi_n$, $\hat{S}(\tau) =
\frac{1}{\beta} \sum_m \eu^{\iu \Omega_m \tau} \hat{S}_m$, $\chi_\qv(\tau-\tau') = \frac{1}{\beta} \sum_m \chi_\qv
(\Omega_m) \eu^{\iu \Omega_m (\tau-\tau')}$ and $\chi_\qv(\Omega_m) = \frac{\chi_0}{\Omega_m^2 + v_s^2 (\qv-\Qv)^2 +
v_s^2 \xi^{-2}}$.

We next insert the identity $\int \mathcal{D} [\vec{S}]\, \eu^{-\frac{\beta}{2} \sum_\qv \chi^{-1}_\qv(\Omega_m)
\vec{S}_{\qv,m} \vec{S}_{-\qv,-m}} = \mathbb{1}$, and through the Hubbard-Stratonovich transformation, the partition
function becomes
\begin{align}
\mathcal{Z} =&  \int \mathcal{D} \left[\bar{\psi}, \psi, \vec{S} \right] \eu^{- \mathcal{S}_{sf}}\non\\
\mathcal{S}_{sf} =& \sum_{\kv, s,n,\kv', s', n'} \bar{\psi}_{\kv, s,n} \left(-(\iu \omega_n - \epsilon_\kv+\mu) \delta_{
\kv \kv'} \delta_{ s s'} \delta_{n n'} - g\, \vec{S}_{\qv, m} \cdot \frac{\vec{\sigma}_{s s'}}{2} \delta_{\kv+\qv,
\kv'} \delta_{m+n, n'} \right) \psi_{\kv', s', n'}\non\\
& \quad\quad+ \frac{\beta}{2} \sum_{\qv, m} \chi_\qv^{-1} (\Omega_m) \vec{S}_{\qv, m}\cdot \vec{S}_{-\qv, -m}
\label{appeq:action0}
\end{align}

Note that no approximation is made to obtain Eq.~\eqref{appeq:action0}, but it cannot be solved exactly in general.

To consider \emph{only} the static spin fluctuations, we restrict to the zero Matsubara frequency for the spin field
$\vec{S}$; the partition function becomes
\begin{align}
\mathcal{Z} =&  \int \mathcal{D} \left[\bar{\psi}, \psi, \vec{S} \right] \eu^{- \mathcal{S}_{\rm static}}\non\\
\mathcal{S}_{\rm static} =& \sum_{\kv, s,n,\kv', s'} \bar{\psi}_{\kv, s,n} \left(-(\iu \omega_n - \epsilon_\kv+\mu) \delta_{
\kv \kv'} \delta_{ s s'}  - 2g\, \vec{S}_{\qv} \cdot \frac{\vec{\sigma}_{s s'}}{2} \delta_{\kv+\qv, \kv'} \right)
\psi_{\kv', s', n}+ \frac{\beta}{2} \sum_{\qv} \chi_\qv^{-1} \vec{S}_{\qv}\cdot \vec{S}_{-\qv}.
\label{appeq:action01}
\end{align}
Here, we have replaced $\vec{S}_{\qv}$ with $\vec{S}_{\qv,0}$ for convenience, and $\chi_\qv^{-1}=(v_s^2 (\qv-\Qv)^2
+v_s^{2}\xi^{-2})/\chi_0$ is the static spin susceptibility. Eq.~\eqref{appeq:action01} may be viewed as an annealed
disorder problem, with $\vec{S}_\qv$ as the static spin impurity. Integrating out the fermion field $\psi$, we get
the effective action in terms of only the spin fields
\begin{align}
\mathcal{Z} = \int \mathcal{D} \left[\vec{S} \right] \exp\left(\tr \ln \mathcal{M}(\vec{S}) - \frac{\beta}{2}
\sum_{\qv} \chi_\qv^{-1} \vec{S}_{\qv}\cdot \vec{S}_{-\qv}\right)
\end{align}
$[\mathcal{M}(\vec{S})]_{\kv s, \kv' s'} = \left((\iu \omega_n - \epsilon_\kv+\mu) \delta_{ \kv \kv'} \delta_{ s s'}  +
g\, \vec{S}_{\kv'-\kv} \cdot \vec{\sigma}_{s s'}\right)$ is the inverse Green's function in a
particular spin configuration determined by $\vec{S}_{\kv'-\kv}$.
To compute the n-point correlation function, we define the generating functional as
\begin{align}
\mathcal{W}(\bar{\eta}, \eta) =  \int \mathcal{D} \left[\bar{\psi}, \psi, \vec{S} \right]\, \exp
\left(-\left(\bar{\psi} \eta + \bar{\eta} \psi + \mathcal{S}_{\rm static}\right)\right),
\end{align}
with the shorthand notation $\bar{\psi} \eta = \sum_{\kv, s , n} \bar{\psi}_{\kv, s ,n} \eta_{\kv, s, n}$.

The full Green's function, i.e. the two-point correlation function, reads
\begin{align}
G(\iu \omega_n, \kv,\kv')_{s s'} & = \langle G(\iu \omega_n, \kv,\kv')  |  \vec{S})_{s s'} \rangle_{\vec{S}} =
\frac{1}{\mathcal{Z}} \frac{\partial^2 \mathcal{W}(\eta, \bar{\eta})}{\partial \bar{\eta}_s \partial
\eta_{s'}}|_{\eta, \bar{\eta}=0} \non\\
& = \frac{ \int \mathcal{D}  \left[\vec{S} \right] [\mathcal{M}(\vec{S})^{-1}]_{\kv s, \kv' s'}\,\exp\left(\tr \ln
\mathcal{M}(\vec{S}) - \frac{\beta}{2} \sum_{\qv} \chi_\qv^{-1} \vec{S}_{\qv}\cdot \vec{S}_{-\qv}\right)}{ \int
\mathcal{D}  \left[\vec{S} \right]  \exp\left(\tr \ln \mathcal{M}(\vec{S}) - \frac{\beta}{2} \sum_{\qv} \chi_\qv^{-1}
\vec{S}_{\qv}\cdot \vec{S}_{-\qv}\right)}\non\\
& \approx \frac{ \int \mathcal{D}  \left[\vec{S} \right] [\mathcal{M}(\vec{S})^{-1}]_{\kv s, \kv' s'}\,\exp\left(
-\frac{\beta}{2} \sum_{\qv} \tilde{\chi}_\qv^{-1} \vec{S}_{\qv}\cdot \vec{S}_{-\qv}\right)}{ \int \mathcal{D}
\left[\vec{S} \right]  \exp\left(- \frac{\beta}{2} \sum_{\qv} \tilde{\chi}_\qv^{-1} \vec{S}_{\qv}\cdot
\vec{S}_{-\qv}\right)}
\label{appeq:twopoint}
\end{align}

From the second to the third line, we assume that the feedback effects on $\vec{S}$ from the fermions, written as
$\tr \ln \mathcal{M}(\vec{S}) $, can be fully captured by replacing the spin susceptibility $\chi$ with a
renormalized one $\tilde{\chi}$, which is determined independently from the four-point correlation function.

The static spin polarization $\Pi^{\alpha\beta}(\qv)$ can be expressed as the four-point correlation function
\begin{align}
\Pi^{\alpha\beta}(\qv) & = \langle \Pi^{\alpha\beta}(\qv  |  \vec{S}) \rangle_{\vec{S}} =\frac{1}{\mathcal{Z}}
\frac{\partial^4 \mathcal{W}(\eta, \bar{\eta})}{\partial \bar{\eta} \partial \eta \partial \bar{\eta} \partial
\eta}|_{\eta, \bar{\eta}=0} \non\\
& \approx \frac{ \int \mathcal{D}  \left[\vec{S} \right] \tr\left[-\frac{1}{2}\sigma^{\alpha}
[\mathcal{M}(\vec{S})^{-1}]_{\kv+\qv , \kv'} \sigma^{\beta} [\mathcal{M}(\vec{S})^{-1}]_{\kv'-\qv , \kv
}\right]\,\exp\left( -\frac{\beta}{2} \sum_{\qv} \tilde{\chi}_\qv^{-1} \vec{S}_{\qv}\cdot \vec{S}_{-\qv}\right)}{
\int \mathcal{D}  \left[\vec{S} \right]  \exp\left( -\frac{\beta}{2} \sum_{\qv} \tilde{\chi}_\qv^{-1}
\vec{S}_{\qv}\cdot \vec{S}_{-\qv}\right)},
\label{appeq:fourpoint}
\end{align}
where the spin index in $\eta, \bar{\eta}$ is omitted.

To determine $\tilde{\chi}_{\qv}$, we note that it is related to the irreducible particle-hole polarization
$\Pi^{\alpha \beta}(\qv)$ as
\begin{align}
\tilde{\chi}_\qv=\frac{\Pi^{\alpha\alpha}(\qv)/2}{1-U \Pi^{\alpha\alpha}(\qv)}
\end{align}
where we have used the fact that due to the SU(2) symmetry in the paramagnetic state, the static spin polarization is
diagonal, i.e.\ $\Pi^{\alpha,\beta\neq \alpha}(\qv)=0$. Assuming that $\chi_\qv$ takes the standard Ornstein-Zernike
form near $\qv\approx \Qv$, i.e.\ $\tilde{\chi}_\qv =\chi_0/( v_s^2 (\qv-\Qv)^2 +v_s^2 \xi^{-2})$, the spin
correlation length $\xi$ in $\tilde{\chi}_\qv$ reads
\begin{align}
\xi^{-2}  = \frac{\chi_0}{v_s^2} \frac{1-U\Pi^{zz}(\Qv)}{\Pi^{zz}(\Qv)}
\approx \frac{2U\chi_0}{v_s^2} (1-U\Pi^{zz}(\Qv))
\label{appeq:xi}
\end{align}
Plugging~\eqref{appeq:fourpoint} into \eqref{appeq:xi}, we can solve for $\xi$ self-consistently. To simplify the
evaluation, it is convenient to integrate out the spin fields $\vec{S}$ and obtain a compact form
for~\eqref{appeq:twopoint} and~\eqref{appeq:fourpoint}. Here, following the suggestion from the one-loop calculation
as demonstrated in Sec.~\ref{sec:summation}, we ignore the spacial fluctuations of the fermion fields. This allows us
to replace $\vec{S}_\qv$ with $\vec{S}_\Qv$ in the fermion propagator, i.e.\
\begin{align}
[\mathcal{M}(\vec{S})]_{\kv s, \kv' s'} \approx \left((\iu \omega_n - \epsilon_\kv) \delta_{ \kv \kv'} \delta_{ s s'}
+ g \beta^{-1} \vec{S}_{\Qv} \cdot \vec{\sigma}_{s s'} \delta_{\kv+\Qv, \kv'} \right).
\label{appeq:M}
\end{align}
Including only the spatial fluctuations for the spin fields, we have
\begin{align}
\exp\left( -\frac{1}{2\beta} \sum_{\qv} \tilde{\chi}_\qv^{-1} \vec{S}_{\qv}\cdot \vec{S}_{-\qv}\right) \approx
\exp\left( -\frac{1}{2\beta} \left(\sum_{\qv} \tilde{\chi}_\qv\right)^{-1} \vec{S}_{\Qv}\cdot \vec{S}_{-\Qv}\right)
=\exp\left(-4\,\bbt^{-1} \vec{S}_\Qv\cdot \vec{S}_\Qv \right),
\label{appeq:chi}
\end{align}
where we remind $\bbt=\frac{4T}{\pi J} \ln \frac{\pi^2/2+\xi^{-2}}{\epsilon^2 + \xi^{-2}}$. To obtain the last line,
we have rescaled $\vec{S}_\Qv$ as $\beta^{-1} \vec{S}_\Qv \Rightarrow \vec{S}_\Qv$, and will use this definition
hereafter. Using Eqs.~\eqref{appeq:M} and~\eqref{appeq:chi}, the two- and four-point correlation functions are
approximated as
\begin{align}
&G(\iu \omega_n, \kv,\kv')_{s s'}  = \langle G(\iu \omega_n, \kv,\kv')  |  \vec{S})_{s s'} \rangle_{\vec{S}} \approx
\frac{ \int \diff  \vec{S}_\Qv \, [\mathcal{M}(\vec{S})^{-1}]_{\kv s, \kv' s'}\,\exp\left(-4\,\bbt^{-1}
\vec{S}_\Qv\cdot \vec{S}_\Qv \right)}{ \int  \diff  \vec{S}_\Qv  \exp\left(-4\,\bbt^{-1} \vec{S}_\Qv\cdot \vec{S}_\Qv
\right)}\non\\
&\Pi^{\alpha\beta}(\qv)  = \langle \Pi^{\alpha\beta}(\qv  |  \vec{S}) \rangle_{\vec{S}} \approx  -\frac{1}{2}\frac{
\int \diff  \vec{S}_\Qv \sum_{s_i}\sigma^{\alpha}_{s_1 s_2} [\mathcal{M}(\vec{S})^{-1}]_{\kv+\Qv s_2, \kv' s_3}
\sigma^{\beta}_{s_3 s_4} [\mathcal{M}(\vec{S})^{-1}]_{\kv'-\Qv s_4, \kv s_1}\, \exp\left(-4\,\bbt^{-1}
\vec{S}_\Qv\cdot \vec{S}_\Qv \right)}{  \int \diff  \vec{S}_\Qv   \exp\left(-4\,\bbt^{-1} \vec{S}_\Qv\cdot
\vec{S}_\Qv \right)}
\label{appeq:result}
\end{align}
where
\begin{align}
 [\mathcal{M}(\vec{S})^{-1}]_{\kv,\kv'} = \frac{1}{1-g^2/4 \vec{S}_\Qv \cdot \vec{S}_\Qv \mathcal{H}_{\kv, n} }
 \begin{pmatrix}
 G^{(0)}(\kv, \iu \omega_n)\mathbb{1}_\sigma \delta_{\kv,\kv'} & -\frac{g}{2} \vec{S}_\Qv \cdot \vec{\sigma}
 \mathcal{H}_{\kv, n} \delta_{\kv+\Qv,\kv'} \\
 -\frac{g}{2} \vec{S}_\Qv \cdot \vec{\sigma} \mathcal{H}_{\kv, n}\delta_{\kv,\kv'+\Qv} &  G^{(0)}(\kv+\Qv, \iu
 \omega_n) \mathbb{1}_\sigma \delta_{\kv+\Qv,\kv'+\Qv}
 \end{pmatrix}
\end{align}
with $\mathcal{H}_{\kv, n} = G^{(0)}(\kv, \iu \omega_n) G^{(0)}(\kv+\Qv, \iu \omega_n) $.

Below and close to $T_N$, we restrict the spin fluctuations to the transverse channel, i.e.\ $\vec{S}_\Qv =\left(
\mathsf{S}_x, \mathsf{S}_y, \langle S_z \rangle \right) = \left( \mathsf{S}_x, \mathsf{S}_y, \Delta/U \right)$, and
only $\mathsf{S}_x, \mathsf{S}_y$ are the static fluctuating fields. Now, we identify the coupling $g/2$ with the
Hubbard interaction $U$. Eq.~\eqref{appeq:result} becomes
\begin{align}
G( \iu \omega_n,\kv,\kv)_{ss}  &= \frac{G^{(0)}(\kv, \iu \omega_n)}{1-\Delta^2 \mathcal{H}_{\kv, n}} \int_0^\infty
\frac{1}{1-u_\omega t}\exp(-t)\non\\
G(\iu \omega_n; \kv, \kv+\Qv)_{ss}  &= \sgn{s}\frac{-\Delta \mathcal{H}_{\kv,\omega}}{1-\Delta^2 \mathcal{H}_{\kv,
n}} \int_0^\infty \frac{1}{1-u_\omega t}\exp(-t)\non\\
\Pi^{zz}(\qv) & {\stackrel{T>T_N}{=}}  \frac{-4}{\bbt}T\sum_{n,\kv} \int \diff \mathsf{S}_x \diff \mathsf{S}_y
\frac{2 \mathcal{H}_\kv}{1-U^2 (\mathsf{S}_x^2 + \mathsf{S}_y^2 )\mathcal{H}_{\kv, n}} \exp (- 4 \bbt^{-1}
(\mathsf{S}_x^2 + \mathsf{S}_y^2 ))\non\\
& = -\frac{1}{\bbt (U/2)^2} \int \frac{\diff \omega}{\pi} n_F (\omega) \im \left[ \int_0^\infty \diff t
\frac{1}{t-u_\omega^{-1}} \exp(-t)\right] \non\\
&= -\frac{1}{\bbt (U/2)^2} \int \diff \omega n_F(\omega) \sgn(\bar{\omega})\Theta(\bar{\omega}-(\emi^{-}_\kv)^2)
\exp\left(-\frac{\bar{\omega}^2 - (\emi^{-}_\kv)^2 }{\bbt (U/2)^2}\right)
\end{align}
where $u_\omega = \frac{(U/2)^2 \mathcal{H}_{\kv, \omega}}{1-\Delta^2 \mathcal{H}_{\kv, \omega}}$,
$n_F(\omega)=\left(\exp(\omega)+1\right)^{-1}$, $\bw=\omega+\mu-\varepsilon^{+}_\kv$,
$E_\kv^2=(\emi^{-}_\kv)^2+\Delta^2$. At $T>T_N$, $\Delta=0$.

\end{document}